\documentclass[%
 reprint,
superscriptaddress,
 amsmath,amssymb,
 aps,
]{revtex4-1}

\usepackage{soul}
\usepackage{graphicx}% Include figure files
\usepackage{dcolumn}% Align table columns on decimal point
\usepackage{bm}% bold math
\usepackage{mathtools}

\usepackage[dvipsnames]{xcolor}

\usepackage[colorlinks=true,citecolor=blue]{hyperref}
\usepackage[normalem]{ulem}
\usepackage[caption=false]{subfig} 
\usepackage{amsmath}
\usepackage{braket}
\usepackage{xspace}

\usepackage{multirow}
\usepackage{capt-of}
\usepackage{float}

\newcommand{\UIB}{Departament de F\'isica, Universitat de les Illes Balears, IAC3 -- IEEC, Crta. Valldemossa km 7.5, E-07122 Palma, Spain}

\newcommand{\AEI}{Max Planck Institute for Gravitational Physics (Albert Einstein Institute), Am Mühlenberg 1, D-14476 Potsdam, Germany}

\newcommand{\mpr}{m^{\prime}}

\newcommand{\phXPHM}{\texttt{IMRPhenomXPHM}}

\newcommand{\phTPHM}{\texttt{IMRPhenomTPHM}}
\newcommand{\phPvthreehm}{\textsc{IMRPhenomPv3HM}\xspace}

\newcommand{\seobnrvforphm}{\textsc{SEOBNRv4PHM}\xspace}

\definecolor{dodgerblue}{HTML}{1E90FF}
\definecolor{viennared}{HTML}{DA0A14}
\definecolor{ctorange}{HTML}{FF6C0C}
\definecolor{granadagreen}{HTML}{078931}
\definecolor{wales}{HTML}{ff0038}
\definecolor{valenciacfred}{HTML}{ee3524}
\definecolor{barcelonafcgold}{HTML}{edbb00}
\definecolor{jam}{HTML}{A50B5E}
\definecolor{austriawien}{HTML}{441678}

\AtBeginDocument{%
  \hypersetup{
    citecolor=dodgerblue,
    linkcolor=dodgerblue,   
    urlcolor=dodgerblue}}

\def\MOneSourceCIPhPHM{\ensuremath{28.1_{-4.3}^{+4.8}}\xspace}
\def\MTwoSourceCIPhPHM{\ensuremath{8.8_{-1.1}^{+1.5}}\xspace}
\def\MtotalSourceCIPhPHM{\ensuremath{36.9_{-2.9}^{+3.7}}\xspace}
\def\ChirpMassSourceCIPhPHM{\ensuremath{13.2_{-0.3}^{+0.5}}\xspace}
\def\MassRatioCIPhPHM{\ensuremath{0.31_{-0.07}^{+0.12}}\xspace}

\def\ChiEffCIPhPHM{\ensuremath{0.22_{-0.11}^{+0.08}}\xspace}
\def\ChiPCIPhPHM{\ensuremath{0.31_{-0.17}^{+0.24}}\xspace}
\def\SpinMagOneCIPhPHM{\ensuremath{0.41_{-0.24}^{+0.22}}\xspace}

\def\DLCIPhPHM{\ensuremath{740_{-190}^{+150}}\xspace}
\def\RedshiftCIPhPHM{\ensuremath{0.15_{-0.04}^{+0.03}}\xspace}
\def\ThetaJNCIPhPHM{\ensuremath{0.71_{-0.27}^{+0.39}}\xspace}
\def\HSNRCIPhPHM{\ensuremath{9.5_{-0.3}^{+0.2}}\xspace}
\def\LSNRCIPhPHM{\ensuremath{16.1_{-0.3}^{+0.2}}\xspace}
\def\VSNRCIPhPHM{\ensuremath{3.6_{-1.0}^{+0.3}}\xspace}
\def\NetSNRCIPhPHM{\ensuremath{19.0_{-0.3}^{+0.2}}\xspace}

\def\MOneSourceCICombined{\ensuremath{29.7_{-5.3}^{+5.0}}\xspace}
\def\MTwoSourceCICombined{\ensuremath{8.4_{-1.0}^{+1.8}}\xspace}
\def\MtotalSourceCICombined{\ensuremath{38.1_{-3.7}^{+4.0}}\xspace}
\def\ChirpMassSourceCICombined{\ensuremath{13.3_{-0.3}^{+0.4}}\xspace}
\def\MassRatioCICombined{\ensuremath{0.28_{-0.06}^{+0.13}}\xspace}

\def\ChiEffCICombined{\ensuremath{0.25_{-0.11}^{+0.08}}\xspace}
\def\ChiPCICombined{\ensuremath{0.30_{-0.15}^{+0.19}}\xspace}
\def\SpinMagOneCICombined{\ensuremath{0.43_{-0.26}^{+0.16}}\xspace}

\def\DLCICombined{\ensuremath{730_{-170}^{+140}}\xspace}
\def\RedshiftCICombined{\ensuremath{0.15_{-0.03}^{+0.03}}\xspace}
\def\ThetaJNCICombined{\ensuremath{0.73_{-0.24}^{+0.34}}\xspace}
\def\HSNRCICombined{\ensuremath{9.5_{-0.3}^{+0.1}}\xspace}
\def\LSNRCICombined{\ensuremath{16.2_{-0.3}^{+0.1}}\xspace}
\def\VSNRCICombined{\ensuremath{3.6_{-1.0}^{+0.3}}\xspace}
\def\NetSNRCICombined{\ensuremath{19.1_{-0.3}^{+0.1}}\xspace}

\def\MOneSourceCIEOBPHM{\ensuremath{31.7_{-3.5}^{+3.6}}\xspace}
\def\MTwoSourceCIEOBPHM{\ensuremath{8.0_{-0.7}^{+0.9}}\xspace}
\def\MtotalSourceCIEOBPHM{\ensuremath{39.7_{-2.7}^{+3.0}}\xspace}
\def\ChirpMassSourceCIEOBPHM{\ensuremath{13.3_{-0.3}^{+0.3}}\xspace}
\def\MassRatioCIEOBPHM{\ensuremath{0.25_{-0.04}^{+0.06}}\xspace}

\def\ChiEffCIEOBPHM{\ensuremath{0.28_{-0.08}^{+0.06}}\xspace}
\def\ChiPCIEOBPHM{\ensuremath{0.31_{-0.15}^{+0.14}}\xspace}
\def\SpinMagOneCIEOBPHM{\ensuremath{0.46_{-0.15}^{+0.12}}\xspace}

\def\DLCIEOBPHM{\ensuremath{740_{-130}^{+120}}\xspace}
\def\RedshiftCIEOBPHM{\ensuremath{0.15_{-0.02}^{+0.02}}\xspace}
\def\ThetaJNCIEOBPHM{\ensuremath{0.71_{-0.21}^{+0.23}}\xspace}
\def\HSNRCIEOBPHM{\ensuremath{9.5_{-0.2}^{+0.1}}\xspace}
\def\LSNRCIEOBPHM{\ensuremath{16.2_{-0.2}^{+0.1}}\xspace}
\def\VSNRCIEOBPHM{\ensuremath{3.7_{-0.5}^{+0.2}}\xspace}
\def\NetSNRCIEOBPHM{\ensuremath{19.1_{-0.2}^{+0.2}}\xspace}

\def\MOneSourceCIXPHM{\ensuremath{30.0^{+5.2}_{-4.3}}\xspace}
\def\MTwoSourceCIXPHM{\ensuremath{8.4^{+1.3}_{-1.1}}\xspace}
\def\MtotalSourceCIXPHM{\ensuremath{38.4^{+4.2}_{-3.2}}\xspace}
\def\ChirpMassSourceCIXPHM{\ensuremath{13.3^{+0.5}_{-0.4}}\xspace}
\def\MassRatioCIXPHM{\ensuremath{0.28^{+0.09}_{-0.07}}\xspace}

\def\ChiEffCIXPHM{\ensuremath{0.25^{+0.1}_{-0.1}}\xspace}
\def\ChiPCIXPHM{\ensuremath{0.23^{+0.20}_{-0.13}}\xspace}
\def\SpinMagOneCIXPHM{\ensuremath{0.39^{+0.16}_{-0.17}}\xspace}

\def\DLCIXPHM{\ensuremath{734^{+161}_{-187}}\xspace}
\def\RedshiftCIXPHM{\ensuremath{0.15_{-0.04}^{+0.03}}\xspace}
\def\ThetaJNCIXPHM{\ensuremath{0.75^{+0.36}_{-0.28}}\xspace}
\def\HSNRCIXPHM{\ensuremath{9.4^{+0.2}_{-0.3}}\xspace}
\def\LSNRCIXPHM{\ensuremath{16.1^{+0.2}_{-0.3}}\xspace}
\def\VSNRCIXPHM{\ensuremath{3.6^{+0.3}_{-0.8}}\xspace}
\def\NetSNRCIXPHM{\ensuremath{18.9_{-0.3}^{+0.2}}\xspace}

\def\MOneSourceCITPHM{\ensuremath{30.9^{+3.5}_{-3.2}}\xspace}
\def\MTwoSourceCITPHM{\ensuremath{8.2^{+0.8}_{-0.7}}\xspace}
\def\MtotalSourceCITPHM{\ensuremath{39.1^{+2.8}_{-2.5}}\xspace}
\def\ChirpMassSourceCITPHM{\ensuremath{13.3^{+0.3}_{-0.3}}\xspace}
\def\MassRatioCITPHM{\ensuremath{0.27^{+0.06}_{-0.05}}\xspace}

\def\ChiEffCITPHM{\ensuremath{0.27^{+0.07}_{-0.07}}\xspace}
\def\ChiPCITPHM{\ensuremath{0.28^{+0.15}_{-0.13}}\xspace}
\def\SpinMagOneCITPHM{\ensuremath{0.44^{+0.14}_{-0.15}}\xspace}

\def\DLCITPHM{\ensuremath{723^{+112}_{-124}}\xspace}
\def\RedshiftCITPHM{\ensuremath{0.15^{+0.02}_{-0.02}}\xspace}
\def\ThetaJNCITPHM{\ensuremath{0.75^{+0.26}_{-0.21}}\xspace}
\def\HSNRCITPHM{\ensuremath{9.5^{+0.1}_{-0.2}}\xspace}
\def\LSNRCITPHM{\ensuremath{16.2^{+0.1}_{-0.2}}\xspace}
\def\VSNRCITPHM{\ensuremath{3.8^{+0.3}_{-0.5}}\xspace}
\def\NetSNRCITPHM{\ensuremath{19.0^{+0.2}_{-0.2}}\xspace}

\begin{document}

%\preprint{APS/123-QED}

\title{New twists in compact binary waveform modelling: a fast time domain model for precession}% Force line breaks with \\

\author{H\'ector Estell\'es}
\affiliation{ \UIB}

%\author{The UIB all-star band}
%\affiliation{ \UIB}

\author{Marta Colleoni}
\affiliation{ \UIB}

\author{Cecilio Garc\'ia-Quir\'os}
\affiliation{ \UIB}

\author{Sascha Husa}
\affiliation{ \UIB}

\author{David Keitel}
\affiliation{ \UIB}

\author{Maite Mateu-Lucena}
\affiliation{ \UIB}

\author{Maria de Lluc Planas}
\affiliation{ \UIB}

\author{Antoni Ramos-Buades}
\affiliation{ \AEI}
\affiliation{ \UIB}

\date{\today}% It is always \today, today,

\begin{abstract}

We present \texttt{IMRPhenomTPHM}, a phenomenological model for the gravitational wave signals emitted by the coalescence of quasi-circular precessing binary black holes systems. The model is based on the ``twisting up'' approximation, which maps non-precessing signals to precessing ones in terms of a time dependent rotation described by three Euler angles, and which has been utilized in several frequency domain waveform models that have become standard tools in gravitational wave data analysis \cite{Hannam:2013oca,Bohe:PPv2,Khan_2019,Khan_2020,phenomxphm}. Our  model is however constructed in the time domain, which allows several improvements over the frequency domain models: we do not use the stationary phase approximation, we employ a simple approximation for the precessing Euler angles for the ringdown signal, and we implement a new method for computing the Euler angles through the evolution of the spin dynamics of the system, which is more accurate and also computationally efficient.

\end{abstract}

\maketitle

%\tableofcontents

\section{Introduction}\label{sec:intro}

A number of complementary strategies have been developed to model the gravitational wave (GW) signal of coalescing compact binaries (CBC), giving rise to several families of waveform models which are routinely used for GW data analysis. Most notably these include two families of effective-one-body (EOB) descriptions, SEOB \cite{Bohe:2016gbl,Cotesta_2018,ossokine2020multipolar}, and TEOBResumS \cite{Nagar_2019_PA,Nagar_2018_TEOB,Nagar_2020}, reduced order methods for parameter space interpolation of numerical relativity (NR) data sets commonly referred to as NRSurrogate models \cite{Blackman_2017,Varma_2019, Varma_2019_prec}, and the IMRPhenom family \cite{Husa:2015iqa,Khan:2015jqa,Hannam:2013oca,Bohe:PPv2,phenomhm,Khan_2019,Khan_2020,pratten2020setting,garcaquirs2020imrphenomxhm,phenomxphm,estells2020time,estells2020imrphenomtp}, which is based on piecewise closed-form phenomenological models, which are particularly computationally efficient due to the closed form expressions.

These models are continuously being improved with the goal to minimize systematic errors when estimating the source parameters of detected GW events with the methods of Bayesian inference \cite{Veitch:2014wba,Ashton_2019}, while reducing at the same time the computational cost of such analyses. 
Such improvements are particularly urgent due to the advances in sensitivity of the international network of advanced GW detectors, and the corresponding increase in the number of detected sources. 
The Advanced LIGO detectors \cite{aLIGO2015} and Advanced Virgo detector \cite{Acernese_2014} have already provided two catalogs of GW transients (GWTC-1 \cite{Abbott_2019} and GWTC-2 \cite{Abbott:2020niy}), which include a total of 50 CBC signals, and a significantly higher number of events is expected for the upcoming O4 observation run \cite{observationscene}.

For the sub-space of quasi-circular binary black hole (BBH) coalescence without spin precession, waveform models have reached a certain level of maturity: The waveform modelling programs mentioned above have all provided models that are calibrated to NR simulations and include several subdominant harmonics of the signal, resulting in very good agreement in the region of parameter space where NR simulations are available, see e.g.~\cite{collaboration2020gw190412,colleoni190412}. Interesting questions remain, e.g. concerning high mass ratios or spins close to the Kerr limit, and models are expected to be further improved, in particular as more high quality NR simulations become available for large mass ratios and spins.
When adding precessing spins such a level of maturity has not yet been reached. Not only is the morphology of waveforms much more complicated, but also the larger parameter space is by far not as well sampled by NR waveforms. Different modelling programs have made different types of compromises: NR surrogate models have been constructed to interpolate NR data sets, but are restricted in coverage of mass ratio, spin magnitudes and length of the waveform \cite{Varma_2019_prec}. The EOB and phenomenological waveform programs have taken a complementary path, using approximations to model precession without calibration to numerical waveforms. This is in principle less accurate, but allows the construction of models that can be used for large parts of the parameter space and without limitations on the length of waveforms. 

The crucial approximation that allows to construct precessing waveform models that are not calibrated to NR is based on the fact that at least during the inspiral the precessing motion is much slower than the orbital motion, so in consequence the precessing motion contributes relatively little to the loss of energy due to gravitational radiation, and therefore contributes relatively little to the phasing of the inspiral. The main effect of precession is then an amplitude modulation as the orbital plane precesses and radiates GWs predominantly in the direction orthogonal to the orbital plane. These arguments can be extended to a drastic simplification of the waveform, by describing it not in an inertial frame, as is appropriate for observation, but rather in a co-precessing non-inertial frame, where the waveform is close to a non-precessing one. One can then construct an approximate precessing waveform in the following way, which is often referred to as ``twisting up'' a non-precessing signal: An appropriate non-precessing waveform is rotated into the inertial frame with a time-dependent rotation, described by three Euler angles (or alternatively quaternions). In order to account for the change of final spin due to precession, the ringdown of the non-precessing waveform also needs to be modified. A standard way to obtain the Euler angles is from post-Newtonian equations, which could be solved directly as time evolution equations, or approximated further in terms of analytic solutions, utilizing orbital averaging \cite{Marsat:2013wwa} or the multiple scale analysis (MSA) \cite{Chatziioannou_2017}.
These ideas have been developed in a series of papers \cite{Schmidt_2011, Boyle_2011, Schmidt_2012, Schmidt_2015, PhysRevD.85.084003} that have resulted in a number of frequency domain precessing IMRPhenom waveform models \cite{Hannam:2013oca,Bohe:PPv2,Khan_2019,Khan_2020,phenomxphm} which have become standard tools in GW data analysis. A recent discussion of the approximations used and their shortcomings has been given in \cite{Ramos-Buades:2020noq}.

For data analysis methods based on matched filtering, it is particularly convenient and computationally efficient to use waveform models in the frequency domain.  In the context of phenomenological waveforms in the frequency domain, the twisting-up approach does however cause several problems: First, in order to obtain closed form expressions for the ``twisted'' spherical harmonic modes in the frequency domain, the stationary phase approximation has been employed, which is not well suited for the merger and ringdown. Second, it has not yet been achieved to obtain a closed form ansatz that computes the Euler angles during ringdown from known information about the quasi-normal modes of the final Kerr black hole, which is straightforward in the time domain \cite{O_Shaughnessy_2013}.
Somewhat surprisingly, it has turned out that precessing IMRPhenom waveform models in the frequency domain are still rather accurate, and they have proven essential tools for GW data analysis. For high-mass systems like GW190521 \cite{Abbott_2020_190521,Abbott_2020_190521_2}, where only the merger and ringdown can be observed, these shortcomings are however essential.

In this work we will therefore treat precession in the time domain. As a starting point we take a non-precessing multimode NR-calibrated model we have constructed recently, \texttt{IMRPhenomTHM} \cite{estells2020time}, and we will generalize it to the precessing \texttt{IMRPhenomTPHM} model employing the ``twisting up'' procedure. We will then discuss the gain in accuracy of describing the merger and ringdown, and new directions of waveform modelling which our approach opens up. A key element is that as an alternative to closed form expressions for the Euler angles we can also apply a fast numerical time integration of the post-Newtonian spin evolution equations, which we hope to develop further in the future.

The paper is organized as follows.
In Sec.~\ref{sec:model} we discuss the model construction, in
Sec.~\ref{sec:results} we evaluate the accuracy and computational efficiency of the model, and we conclude in Sec.~\ref{sec:conclusions}.

\section{Model construction}\label{sec:model}

\subsection{Notation and conventions}\label{sec:notation}

Quasi-circular BBH systems can be described by eight intrinsic parameters: the individual masses $m_i$ and individual dimensionless spin vectors $\boldsymbol{\chi}_i=\boldsymbol{S}_i/m_i^2$ of each black hole component. The total mass of the system $M=m_1+m_2$ is a scale parameter and can be used to define geometric units $G=c=M=1$. We define the mass ratio $q=m_2/m_1 \geq 1$ and the symmetric mass ratio $\eta=q/(1+q)^2$.
We denote dimensionless component spin vectors by $\boldsymbol{\chi}_i$, dimensionful spins by $\boldsymbol{S}_i = m_i^2 \, \boldsymbol{\chi}_i$ and the orbital angular momentum by ${\boldsymbol{L}}$.

The emitted GW signal in a direction $(\Theta,\Phi)$ on the celestial sphere of the source can be expressed in a polarization basis in terms of two independent polarizations, or decomposed in a basis of spherical harmonics of spin weight $-2$:
\begin{equation}
\begin{split}
    h(t;\boldsymbol{\lambda},\Theta,\Phi) & =h_{+}(t;\boldsymbol{\lambda},\Theta,\Phi) - ih_{\times}(t;\boldsymbol{\lambda},\Theta,\Phi)\\
    &=\sum_l\sum_{m=-l}^{l}h_{lm}(t;\boldsymbol{\lambda}) \, {}^{-2}Y_{lm}(\Theta,\Phi),
\end{split}
\end{equation}
which disentangles the extrinsic orientation parameters $(\Theta,\Phi)$ from the intrinsic parameters $\boldsymbol{\lambda}=\{q,\boldsymbol{\chi}_1,\boldsymbol{\chi}_2\}$.

For non-precessing systems, the spherical harmonic modes are naturally defined with respect to an axis that is orthogonal to the preserved orbital plane. In the presence of spin precession the orbital plane also precesses. There is then no natural definition of a frame in which to define the spherical harmonic modes, and indeed the complexity of their morphology depends greatly on the definition of the (inertial) frame.

A particularly simple form of the GW signal can be achieved when the axis of the inertial reference frame is aligned with the total angular momentum of the system
\begin{equation}
    \boldsymbol{J}(t)=\boldsymbol{L}(t) + \boldsymbol{S}_1(t) + \boldsymbol{S}_2(t),
\end{equation}
at some given reference time $\hat{\boldsymbol{z}}=\hat{\boldsymbol{J}}(t_\mathrm{ref})$. We will refer to such a reference frame as the $\boldsymbol{J}$-frame. The direction of ${\boldsymbol{J}}$ is approximately constant, but under certain conditions can flip, which is known as ``transitional precession'' \cite{Apostolatos:1994mx}.

Another type of frame that is commonly employed chooses the orbital angular momentum ${\boldsymbol{L}}$ as the axis. 
We can consider ${\boldsymbol{L}}$ as a time-dependent quantity, giving rise to a non-inertial frame, or define an inertial frame in terms of a reference time, where ${\boldsymbol{L}}(t_\mathrm{ref})={\boldsymbol{L_0}}$.
We will refer to these choices as the ${\boldsymbol{L}}$-frame or ${\boldsymbol{L_0}}$-frame. The ${\boldsymbol{L}}$-frame has the advantage that the spin components parallel and orthogonal to ${\boldsymbol{L}}$ are approximately preserved, see e.g. the discussion in \cite{Schmidt_2015}. For this reason the frames associated to the orbital angular momentum are preferred when defining the spin vectors for initial data sets in NR -- the spins will then typically change little over time. Due to the precession of ${\boldsymbol{L}}$ around the total angular momentum ${\boldsymbol{J}}$ this simplicity is however paid for in a more complex morphology of the spherical harmonic modes in the $\boldsymbol{L_0}$ frame.

An alternative to tying the frame axis to ${\boldsymbol{L}}$ is to use instead the Newtonian angular momentum ${\boldsymbol{L}}_\mathrm{N}$, which points in the direction of the instantaneous angular frequency vector. We will implement our ``twisting up'' approach to modelling precessing by mapping the preserved axis of the non-precessing system to the time-dependent ${\boldsymbol{L}}_\mathrm{N}$ of the precessing system.

At some reference time $t_\mathrm{ref}$ we then define a Cartesian coordinate system with the axes
\begin{equation}\label{eq:Lframe}
\hat{\boldsymbol{z}}=\hat{\boldsymbol{L}}_\mathrm{N}(t_\mathrm{ref}), \qquad \hat{\boldsymbol{x}}=\boldsymbol{r_1}(t_\mathrm{ref})-\boldsymbol{r_2}(t_\mathrm{ref}), 
\end{equation}
 as the direction between the larger and the smaller black hole at that reference time, and $\hat{\boldsymbol{y}}$ constructed orthonormally to the other two following the right-hand-side rule. In this frame, the angles that specify the position in the source sky-sphere are commonly called
inclination $\iota$ and reference orbital phase $\phi_\mathrm{ref}$.
In order to simplify our notation we will refer to this frame  as $\boldsymbol{L}_0$-frame, neglecting the differences between ${\boldsymbol{L}}_\mathrm{N}$ and ${\boldsymbol{L}}$.

\subsection{``Twisting-up'' approximation}\label{sec:twisting}

We will construct our precessing waveform models in terms of rotating the spherical harmonic modes of non-precessing systems into the inertial precessing systems as discussed above.
This time-dependent rotation can be characterized by three Euler angles $(\alpha,\beta,\gamma)$ describing the 3D rotation between both frames, and then the relation between the modes can be expressed as \cite{Goldberg:1966uu}:
\begin{equation}
    h^{I}_{lm}(t)=\mathcal{D}^{l}_{mm'}(\alpha,\beta,\gamma)h^\mathrm{cop}_{lm'}(t)\,
\end{equation}
where $\mathcal{D}^{\ell}_{m \mpr}$ are the Wigner D-matrices. Our conventions for the Euler angles and D-matrices are consistent with those used in \cite{phenomxphm}, where we also provide further details on the D-matrices.

The Euler angles can be expressed in the inertial reference frame $\{\hat{\boldsymbol{x}},\hat{\boldsymbol{y}},\hat{\boldsymbol{z}}\}$ defined by eq.~(\ref{eq:Lframe}) as:
\begin{subequations}
\label{eq:eulerangles}
\begin{equation}
    \alpha=\arctan(\hat{L}_y/\hat{L}_x),
\end{equation}
\begin{equation}
    \cos\beta=\hat{\boldsymbol{z}}\cdot\hat{\boldsymbol{L}}=\hat{L}_z,
\end{equation}
\begin{equation}
\label{eq:mrc}
    \dot{\gamma}=-\dot{\alpha}\cos\beta\,.
\end{equation}
\end{subequations}

As discussed in Sec.~\ref{sec:intro} we will approximate the spherical harmonic modes of the precessing waveform in a co-precessing frame by the modes of the corresponding non-precessing (``AS'' for ``aligned spin'') system:
\begin{equation}
    h^{\text{coprec}}_{lm}(t;q,\boldsymbol{\chi}_1,\boldsymbol{\chi}_2)\approx h^{\text{AS}}_{lm}(t;q,\chi_{1l},\chi_{2l}).
\end{equation}
Here we will employ this approximation to map the non-precessing modes from the NR-calibrated model IMRPhenomTHM~\cite{estells2020time} into the precessing inertial modes of the IMRPhenomTPHM model, and we set
\begin{equation}\label{eq:TPHM_from_THM}
    h^\mathrm{TPHM}_{lm}(t) = \mathcal{D}^{l}_{mm'}(\alpha,\beta,\gamma)h^\mathrm{THM}_{lm'}(t)\, .
\end{equation}

In our construction, the spherical harmonic modes resulting from eq.~(\ref{eq:TPHM_from_THM}) are given in the inertial $\boldsymbol{J}$-frame, since in this frame the morphology of the Euler angles is simplified. In order to construct the GW polarization time series, which are the measurable quantities at the detectors, we first rotate the spherical harmonic modes to the $\boldsymbol{L}_0$-frame with a global time-independent rotation,
\begin{equation}
    h^\mathrm{\boldsymbol{L}_0}_{lm}(t) = \mathcal{D}^{l}_{mm'}(-\gamma_\mathrm{ref},-\beta_\mathrm{ref},-\alpha_\mathrm{ref})h^\mathrm{\boldsymbol{J}}_{lm'}(t)\, 
\end{equation}
since in this frame the polarizations can be constructed as
\begin{equation}
\label{eq:lalpolariz}
    h_{+}(t) - ih_{\times}(t) = \sum_l\sum_{m=-l}^{l}h^\mathrm{\boldsymbol{L}_0}_{lm}(t) \, {}^{-2}Y_{lm}(\iota,\phi_{\mathrm{ref}}).
\end{equation}

\subsection{Co-precessing modes}\label{sec:coprec}

The \texttt{IMRPhenomTHM} model employed in eq.~(\ref{eq:TPHM_from_THM}) is a non-precessing phenomenological multimode model \cite{estells2020time} that has been calibrated to 531 non-precessing BBH NR simulations. 
In our approximation, as discussed above, we neglect the contribution of the spin components within the orbital plane to the frequency evolution, likewise the evolution of the projection $\chi_{iL}=\boldsymbol{\chi}_{i}(t)\cdot\hat{\boldsymbol{L}}(t)$ is not taken into account for describing the frequency evolution of the precessing system. This approximation has been used in all models from the IMRPhenom family to date. See \cite{Ramos-Buades:2020noq} for a further discussion of the approximations used in the ``twisting'' approximation.

The GW multipoles $h_{\ell m}$ are also described by the \texttt{IMRPhenomTHM} model as piece-wise expressions defined on a three-region partition of the time domain: inspiral, merger and ringdown, as described in \cite{estells2020time}, and similar to the decomposition used in the construction of the \texttt{IMRPhenomX} frequency-domain 
models \cite{pratten2020setting,phenomxphm}. We then modify the ringdown region to account for the final spin of the actual precessing system, which differs in general from the non-precessing case, see our detailed discussion in \cite{phenomxphm}. The ringdown ansatz of  \texttt{IMRPhenomTHM} is based on the phenomenological analytical proposal of \cite{damour2014new}, which employs the quasinormal mode frequencies of the remnant black hole. In \texttt{IMRPhenomTHM} we describe the mode frequencies and amplitudes as
\begin{equation}
\label{eq:rdomega}
\begin{split}
    \bar{\omega}_{lm}(t)&=\omega_{lm}(t) - \omega^\mathrm{RD}_{1lm}\\
    &=c_1\dfrac{c_2(c_3e^{-c_2t} + 2c_4e^{-2c_2t})}{1+c_3e^{-c_2t} + c_4e^{-2c_2\ t}},
\end{split}
\end{equation}
\begin{equation}
\label{eq:rdamp}
\begin{split}
    |\bar{h}_{lm}(t)|&=e^{\alpha_{1lm}(t-t^\mathrm{peak}_{lm})}|h_{lm}|\\
    &=d_1\tanh[d_2 (t-t^\mathrm{peak}_{lm})+d_3] + d_4\,.
\end{split}
\end{equation}
These expressions depend explicitly on the ground-state damping frequency $\alpha_{1lm}$ and ringdown frequency $\omega^\mathrm{RD}_{1lm}$ of each mode, and also some of the coefficients (see equations 31-32 of \cite{estells2020time}) depend on the damping frequency of the first overtone $\alpha_{2lm}$. Damping and ringdown frequencies are functions of the final spin of the remnant black hole, so for the precessing situation we evaluate them with our prediction of the precessing final spin, which we discuss below at the end of Sec.~\ref{sec:EulerAnglesRD}.

\subsection{Description of inspiral-plunge precessing Euler angles} \label{sec:EulerAnglesInsp}

For \texttt{IMRPhenomXPHM}, the corresponding frequency-domain model, two alternative closed-form expressions for the Euler angles have been used \cite{phenomxphm}. Here we inherit these prescriptions, but we add an additional option, to numerically evolve the spin precession equations.

\subsubsection{Numerical evolution of the spin precession equations}\label{sec:Numerical angles}

When the individual spins $\boldsymbol{S}_{1,2}$ are misaligned with respect to the orbital angular momentum $\boldsymbol{L}$, both the spins and the orbital angular momenta directions evolve in time, producing a precessional motion of the orbital plane. Under the assumption of conserved direction of the total angular momentum $\boldsymbol{J}$, and neglecting radiation reaction, the evolution equations for the precessing spins and the orbital angular momentum can be obtained in Post-Newtonian (PN) theory to a given order in the PN expansion parameter $v$:
\begin{subequations}
\label{eq:preceqs}
\begin{align}
    \dfrac{d\hat{\boldsymbol{L}}}{dt}&=\boldsymbol{\Omega}_{\hat{L}}(v(t),q,\boldsymbol{S}_1,\boldsymbol{S}_2)\times\hat{\boldsymbol{L}},\\
    \dfrac{d\boldsymbol{S}_1}{dt}&=\boldsymbol{\Omega}_1(v(t),q,\boldsymbol{S}_1,\boldsymbol{S}_2)\times\boldsymbol{S}_1,\\
    \dfrac{d\boldsymbol{S}_2}{dt}&=\boldsymbol{\Omega}_2(v(t),q,\boldsymbol{S}_1,\boldsymbol{S}_2)\times\boldsymbol{S}_2.
\end{align}
\end{subequations}
Note that these equations are subject to the constraint:
\begin{equation}
    \dot{\boldsymbol{L}}=-\dot{\boldsymbol{S_1}}-\dot{\boldsymbol{S_2}}.
\end{equation}

Radiation reaction can be introduced by letting the PN parameter $v(t)=(\dot{\phi}_\mathrm{orb}(t))^{1/3}$  evolve in time, which implies another ordinary differential equation (ODE) to be solved for $\dot{v}$, which in general will depend on the time-dependent individual spin vectors. In this work however, we approximate the spin evolution by inheriting the evolution of $v(t)$ from the non-precessing analytical orbital frequency from the \texttt{IMRPhenomT} model \cite{estells2020imrphenomtp,estells2020time}, which is defined as half of the wave frequency of the $l=2$, $m=2$ non-precessing mode:
\begin{equation}
\label{eq:phenTorbfreq}
    \phi^{T}_\mathrm{orb}(t)=\phi^{T}_{22}(t)/2.
\end{equation}
This has several advantages: first, the description of the orbital frequency of \texttt{IMRPhenomT} has been calibrated against a set of 531 NR waveforms in the late inspiral and merger along with 63 intermediate-mass-ratio waveforms from the adiabatic solution of the Teukolsky equation.  Thus, while we neglect precession effects as is consistent with our ``twisting-up'' approximation,  $v(t)$ remains regular up to the coalescence time across parameter space. Second, this avoids solving an extra ODE for $\dot{v}(t)$, which is typically the most expensive ODE in the system, accelerating substantially the numerical integration of our system of ODEs. 
In the future we plan to revisit this construction and investigate how to best incorporate information 
about the evolution of the individual spins and about the spin projections perpendicular to the orbital angular momentum.

For the implementation of the PN spin evolution equations (\ref{eq:preceqs}) we rely on the \texttt{SpinTaylor} insfrastructure \cite{SturaniT4} in the \texttt{LALSimulation} module of the \texttt{LALSuite} framework for GW data analysis \cite{lalsuite}. This implementation includes corrections in the spin equations up to next-to-next-to-next-to-next-to-leading order ($\text{N}^4$LO), including both instantaneous terms and orbit averaged terms. The implementation of the evolution equations in \cite{SturaniT4} returns the evolution equation for the Newtonian orbital angular momentum direction $\hat{\boldsymbol{L}}_{\mathrm{N}}(t)$, which up to next-to-leading order NLO agrees with $\hat{\boldsymbol{L}}(t)$. However at higher order, linear terms in $\boldsymbol{L}$ contaminate $\hat{\boldsymbol{L}}(t)$, causing the directions to differ. Although physically it would be desirable to enable these terms, in the implementation of this model is was decided to disable them, since technical problems were observed in the rotated modes when enabling these contributions. We will revisit this decision after further investigation in the future.

Solving the equations (\ref{eq:preceqs}) we obtain the evolution of $\hat{L}(t)$, and from eq.~(\ref{eq:eulerangles}) the Euler angles defined in eqs.~(\ref{eq:eulerangles}), which ``twist'' the co-precessing modes into the inertial precessing modes.

\subsubsection{Analytical expressions: Next-to-next-to-leading order effective single-spin and double-spin multiscale analysis }\label{sec:NNLO}

Besides the default implementation of the precessing Euler angles discussed in the previous subsection, the model inherits the analytical options included in \texttt{IMRPhenomXP} and \texttt{IMRPhenomXPHM}: the next-to-next-to-leading order (NNLO) effective single-spin approximation \cite{Marsat:2013wwa,Boh__2013} and the double-spin multiscale analysis (MSA) approximation \cite{Chatziioannou_2017}. We briefly outline here the main features of both descriptions.

The NNLO effective single-spin description is based on the introduction of the triad $\{\boldsymbol{n},\boldsymbol{\lambda},\boldsymbol{\ell}\}$, where $\boldsymbol{n}$ is the unit separation vector between both black holes, $\boldsymbol{\ell}$ is the direction of the unit vector normal to the instantaneous orbital plane and $\boldsymbol{\lambda}$ completes the triad following the right hand rule $\boldsymbol{\lambda}=\boldsymbol{\ell}\times \boldsymbol{n}$.
The evolution equations for the Euler angles in the single-spin case are:
\begin{subequations}
\begin{align}
    \frac{d \alpha}{d t} &= -\frac{\bar{\omega}}{\sin \beta} \frac{J_n}{\sqrt{J^2_n + J^2_{\lambda}}} , \\
    \frac{d \beta}{d t} &= \bar{\omega} \frac{J_{\lambda}}{\sqrt{J^2_n + J^2_{\lambda}}},
\end{align}
\end{subequations}
where $J_{n,\lambda}$ are the components of the total angular momentum $\boldsymbol{J}=\boldsymbol{L} + \boldsymbol{S}_1$ in this triad. These equations can be solved analytically to next-to-next-to-leading order in the spin-orbit coupling \cite{Boh__2013} (and then $\gamma$ could be obtained analytically from eq. \ref{eq:mrc}), assuming a single spinning black hole, where the spin degrees of freedom of the full problem are mapped to an effective single spin aligned with the direction of the orbital angular momentum
\begin{equation}
    \chi_\mathrm{eff}=\frac{m_1\chi_{1,L} + m_2\chi_{2,L}}{m_1+m_2},
\end{equation}
and a precessing spin parameter 
\begin{equation}
    \chi_\mathrm{p} = \frac{1}{A_1 m_1^2}\max(A_1S_1^{\perp},A_2S_2^{\perp}),
\end{equation}
where $A_1=(2+2/3q)$ and $A_2=(2+3q/2)$, that captures the averaged total in-plane spin and is an approximately conserved quantity. This description was employed in the IMRPhenomP and IMRPhenomPv2 models \cite{Schmidt_2015, Bohe:PPv2} and it is also available in the IMRPhenomXP and IMRPhenomXPHM models as an alternative option \cite{phenomxphm}.

The MSA double-spin description exploits the natural separation among different timescales intervening in the evolution of a precessing binary systems: the orbital timescale $t_\mathrm{orb}$, the precession timescale $t_\mathrm{prec}$ and the radiation reaction timescale $t_\mathrm{RR}$, which during most of the evolution satisfy
\begin{equation}
    t_\mathrm{orb} \ll t_\mathrm{prec} \ll t_\mathrm{RR} \,.
\end{equation}
Based on this separation of timescales, radiation reaction effects can be included as a perturbation on a closed-form solution for the conservative case \cite{Kesden_2015}, where the precessing angle $\alpha(t)$ can be split into two contributions:
\begin{equation}
    \alpha(t) = \alpha_{-1}(t)+\alpha_0(t),
\end{equation}
where $\alpha_{-1}(t)$ is the leading-order term averaged over the precession timescale (which is fast compared with the radiation-reaction timescale) and then integrated over radiation reaction, while the term $\alpha_{0}(t)$ is a first order correction that includes information about the relative orientation between the individual spins (and hence it incorporates double-spin information).

The implementation of these analytical approximations into the model relies on the existing infrastructure for the \texttt{IMRPhenomXP} and \texttt{IMRPhenomXPHM} models \cite{phenomxphm}, although there is an important difference: for the present model, the analytical expressions for the angles are evaluated employing the non-precessing analytical orbital frequency of \texttt{IMRPhenomT} (see eq. (\ref{eq:phenTorbfreq}) and related discussion in previous section), instead of the Fourier frequency as in the Fourier domain models. This, together with the fact that the model construction is native in the time domain, allows us to dispense with the stationary phase approximation (SPA) employed for twisting up the modes in the Fourier domain models, which can enhance the validity of the description in the strong field regime. 

\begin{figure}[htpb!]
\includegraphics[width=\columnwidth]{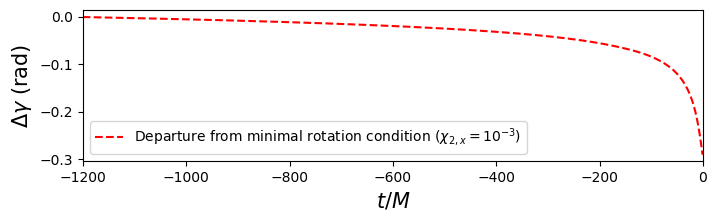}
\includegraphics[width=\columnwidth]{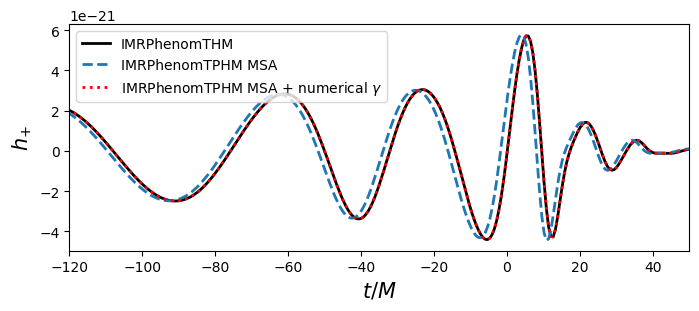}
\includegraphics[width=\columnwidth]{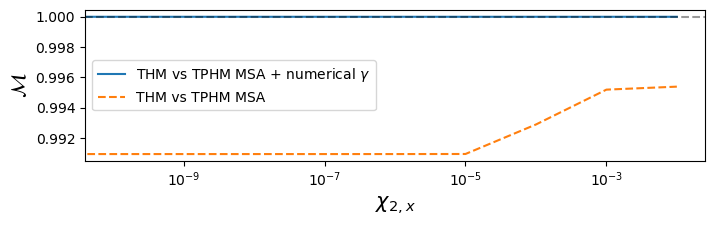}
 \caption{Recovery of the non-precessing limit with analytical MSA description of the Euler angles for the BBH configuration $q=3$, $\chi_{1,z}$=-0.5, $\chi_{2,x}=10^{-3}$ (first two pannels). Top: departure of the analytical third Euler angle from the minimal rotation condition. Middle: Plus-polarization amplitude for the non-precessing model \texttt{IMRPhenomTHM}, for the precessing extension with full analytical MSA, and when computing the third Euler angle from the minimal rotation condition. Bottom: Match recovery of the non-precessing model for small perturbations of the non-precessing case.}
\label{fig:ASlimitMSA}
\end{figure}

We introduce several additional options for the treatment of the NNLO and MSA analytical approximations. First, our implementation allows to replace the analytical evaluation of the third Euler angle $\gamma(t)$ by the numerical evaluation of the minimal rotation condition (eq. \ref{eq:mrc}), once the other two Euler angles $\alpha(t)$ and $\beta(t)$ have been computed.
This is motivated by the realization that the MSA approximation does not correctly satisfy the aligned-spin limit, i.e. models twisted up with this description do not reproduce the underlying non-precessing model in the limit of vanishing in-plane spins. We traced this problem to the fact that the expression for the third Euler angle, which is obtained by applying the MSA expansion to the minimal rotation condition, does not satisfy this condition in this regime. (And probably it does not satisfy it in general.) With this addition, the model is able to reproduce the underlying non-precessing model in the aligned-spin limit. See Fig.\ref{fig:ASlimitMSA} for an illustrative example.

Second, both the NNLO and the MSA approximations lose accuracy before merger due to the breakdown of the underlying PN approximation. In particular, the merger time predicted by the PN approximations will generally be different from the merger time predicted by the non-precessing model, creating some tension.
In order to smooth the behavior in the strong-field regime, we implement an option for substituting the angle description by a linear continuation from the minimum energy circular orbit (MECO) time (which sets the boundary of validity of the underlying adiabatic approximation of the PN Taylor approximants \cite{Cabero_2017}) to the peak time:
\begin{subequations}
\label{eq:linearmerger}
\begin{align}
    \alpha^\mathrm{coal}(t)&=\alpha(t_\mathrm{MECO}) + t\dot{\alpha}(t_\mathrm{MECO}),\\
    \beta^\mathrm{coal}(t)&=\beta(t_\mathrm{MECO}) + t\dot{\beta}(t_\mathrm{MECO}).
\end{align}
\end{subequations}
It is evident that this treatment does not contain actual physical information about the behavior during the strong field regime. But in any case, the complicated angle morphology in this region does not correctly represent the behavior, so we substitute our ignorance about the behavior in this regime by a simpler well-behaved description. The actual usage of these options in the code is explained in Appendix \ref{appen:useropts}.

\subsection{Merger-ringdown treatment of precessing Euler angles}\label{sec:EulerAnglesRD}

\begin{figure}[thpb!]
\includegraphics[width=\columnwidth]{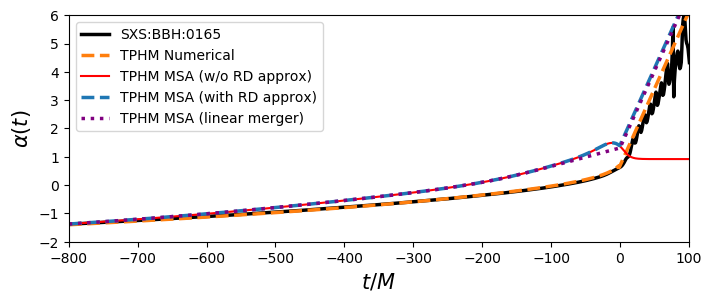}
\includegraphics[width=\columnwidth]{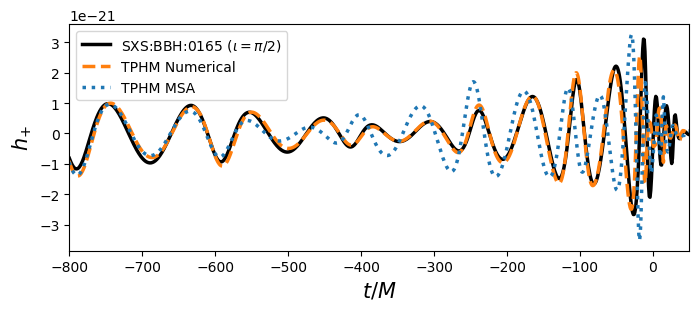}
 \caption{Comparison of the different Euler angle implementations with a challenging NR simulation, \texttt{SXS:BBH:0165} (the worst case in the study detailed below in Sec. \ref{sec:nrcomparison}). Top: comparison of the precessing angle $\alpha(t)$. Bottom: comparison of the plus polarization for edge-on inclination.}
\label{fig:anglecomparison}
\end{figure}

The previous descriptions of the precessing Euler angles apply up to the coalescence time, which we determine according to the \texttt{IMRPhenomT} model as the peak time of the $l=2$, $m=2$ dominant mode. After this time the orbital frequency is not well defined, as the orbiting binary components have merged into a remnant black hole. For the post-merger description of the Euler angles, we rely, as discussed in \cite{estells2020imrphenomtp}, on the realization that the ringdown signal exhibits an effective precessional motion \cite{O_Shaughnessy_2013}. An analytical approximation to this behavior can be derived in the limit of small opening angles, taking the leading contribution of the twisting-up formula considering only the twisting of the $l=2,|m|=2$ co-precessing modes:
\begin{equation}
    h^{P}_{2m}\simeq e^{-im\alpha}e^{-i2\gamma}d^2_{2m}(\beta)h^{\text{coprec}}_{22}\,.
\end{equation}
We then compute the complex ratio between the inertial $m=2$ and $m=1$ modes as
\begin{equation}
\label{eq:ratiord}
    h^{P}_{22}/h^{P}_{21}\simeq -\frac{1}{2}e^{-i\alpha}\tan(\beta/2).
\end{equation}
Expressing the modes in the ringdown as a superposition of QNM states and considering only the leading ground state we obtain
\begin{equation}
    h_{2m}^{\text{RD}}\simeq H_0 e^{-\omega^{\text{damp}}_{12m}}e^{i\omega^{\text{RD}}_{12m}}\,.
\end{equation}
Employing eq. (\ref{eq:ratiord}), the leading contribution to the Euler angles $\alpha$ and $\beta$ (and then $\gamma$ employing the minimal rotation condition) during the ringdown then becomes
\begin{subequations}
\label{eq:anglesrd}
\begin{align}
    \alpha^{\text{RD}}(t)&\simeq (\omega^{\text{RD}}_{122}-\omega^{\text{RD}}_{121})t + \alpha_0^{\text{RD}},\\
    \beta^{\text{RD}}(t)&\simeq -2\arctan\Big(2e^{(\omega^{\text{damp}}_{121}-\omega^{\text{damp}}_{122})t}\Big)+\beta_0^{\text{RD}}.
\end{align}
\end{subequations}
While the expression for $\alpha^{\text{RD}}(t)$  qualitatively reproduces the NR behavior even for situations where the opening angle cannot be considered small (see for example Fig. \ref{fig:anglecomparison} top panel), further investigation to understand the caveats of $\beta^{\text{RD}}(t)$ is needed. For the current implementation of the model it was thus decided to implement only the constant value at the merger time. This will be revisited in future upgrades of the model.

For evaluating the quasinormal frequencies $\omega_{nlm}^{\mathrm{RD}/\mathrm{damp}}$ employed in the previous expression, and also in the construction of the co-precessing ringdown modes amplitude from eq. (\ref{eq:rdamp}) and frequency from eq. (\ref{eq:rdomega}), a prediction for the final spin $\chi_f=S_\mathrm{f}/M_\mathrm{f}^2$ of the remnant black hole is needed. In this model we approximate the final spin of the remnant black hole of the precessing system applying a simple augmentation of the non-precessing final spin based on vector addition of the orbital angular momentum and the component spins, and using approximations consistent with our twisting procedure, as in done in other models like \texttt{IMRPhenomXP} and \texttt{IMRPhenomXPHM}. As discussed in detail in Sec. IV.D of \cite{phenomxphm} we approximate the magnitude of the final spin as
\begin{equation}\label{eq:finalspin}
    \chi_\mathrm{f}^{\text{P,augmented}}=\sqrt{\chi_{\mathrm{f},\mathrm{AS}}^2 + S_{\perp}^2/M^4},
\end{equation}
where $\chi_{\mathrm{f},\mathrm{AS}}$ corresponds to the dimensionless final spin of the equivalent non-precessing configuration (the one employed in the evaluation of the non-precessing modes for approximating the co-precessing modes) and $S_{\perp}$ measures the in-plane spin contribution. For the non-precessing final spin, we employ the same formula as in the non-precessing model \texttt{IMRPhenomTHM}, based on a hierarchical data-driven fit of available NR data \cite{Jim_nez_Forteza_2017}. For the estimation of $S_{\perp}$, we allow the same options discussed in \cite{phenomxphm}, but the default description is based on the evaluation of the evolved individual spins at the merger time, from the evolution of eq. (\ref{eq:preceqs}):
\begin{equation}
    S_{\perp} = \sqrt{|\boldsymbol{S}_{1,\perp}(t_\mathrm{merger})+\boldsymbol{S}_{2,\perp}(t_\mathrm{merger})|^2},
\end{equation}
where $\boldsymbol{S}_{i,\perp}(t_\mathrm{merger})$ are the projections of the individual spins perpendicular to $\hat{\boldsymbol{L}}$ at the coalescence time. Also, in this option the non-precessing final spin fit is evaluated employing the parallel spin components to $\hat{\boldsymbol{L}}$ from the evolved spins at the coalescence time.

Eq.~(\ref{eq:finalspin}) only predicts the magnitude of the final spin. In order to define the direction of the final spin we proceed again as discussed in Sec.~IV.D in \cite{phenomxphm}, tracking the direction of the orbital angular momentum with respect to the orbital plane.

\section{Model performance}\label{sec:results}

In this section we validate the accuracy and computational efficiency of the model. We compare with NR waveforms and other state-of-the-art precessing multimode waveform models based on the mismatch between waveforms, recover the parameters of injected synthetic signals corresponding to NR simulations, and apply Bayesian inference to observed GW events and compare with results for these events from the literature.

As discussed in Appendix \ref{appen:useropts}, the LALSuite \cite{lalsuite} implementation of our model supports several options regarding the choice of approximation for the Euler angle prescription and for the final spin approximation. These options are selected with parameters that take integer values, which we will refer to as $\texttt{PV}$ for precession version and $\texttt{FS}$ for final spin.

\subsection{Comparison with Numerical Relativity}\label{sec:nrcomparison}

\subsubsection{Mismatch comparison with \texttt{LVCNR} catalog}

For checking the agreement between the model and NR waveforms, which are the best source of information that we have about coalescing BBH signals in the strong-field regime, we follow the standard practice of computing the mismatch between waveforms. Taking the standard definition of the inner product in the space of waveforms (see e.g.~\cite{Cutler:1994ys}),
\begin{equation}
\label{eq:nwip}
\Braket{h_1, h_2} = 4 \Re \int_{f_{\min}}^{f_{\max}} \frac{\widetilde{h}_1(f) \:\widetilde{h}^*_2(f)}{S_\mathrm{n}(f)},
\end{equation}
where $S_\mathrm{n}(f)$ is the one-sided power spectral-density (PSD) of the detector noise, the \textit{match} $\mathcal{M}(h_1,h_2)$ is defined as the normalized inner product maximized over relative time and phase shifts between the given set of waveforms:
\begin{equation}
\mathcal{M}(h_1,h_2) = \max_{t_0, \phi_0} \frac{\Braket{h_1, h_2}}{\sqrt{\Braket{h_1, h_1}}\sqrt{\Braket{h_2, h_2}}}.
\end{equation}
The mismatch $\mathcal{MM}(h_1,h_2)$ is defined as the deviation of the match from unity,
\begin{equation}
\mathcal{MM}(h_1,h_2)=1-\mathcal{M}(h_1,h_2).
\end{equation}
For the results presented in this paper, we employ the Zero-Detuned-High-Power PSD \cite{adligopsd}, which models the advanced LIGO~\cite{aLIGO2015} design sensitivity.  

As done in \cite{phenomxphm}, we analytically optimize over the template polarization angle, following \cite{PhysRevD.94.024012}, and numerically optimize over reference phase and rigid rotations of the in-plane spins at the reference frequency. In order to perform the numerical optimization we use the dual annealing algorithm as implemented in the {\tt SciPy} {\tt Python} package \cite{2020SciPy-NMeth}. We then compute the SNR-weighted match $\mathcal{M}_{\mathrm{w}}$ ~\cite{Harry:2016aa}
\begin{equation}
\mathcal{M}_{\mathrm{w}}=\left(\frac{\sum_i\mathcal{M}_i^3 \Braket{h_{i,\mathrm{NR}}, h_{i,\mathrm{NR}}}^{3/2}}{\sum_i{\Braket{h_{i,\mathrm{NR}},h_{i,\mathrm{NR}}}^{3/2}}}\right)^{1/3},
\end{equation}
where the subscript $i$ refers to different choices of polarization and reference phase of the source.

We have computed mismatches for \texttt{IMRPhenomTPHM} against 99 precessing SXS waveforms~\cite{Boyle:2019kee,SXS:catalog}, picking for each binary configuration the highest resolution available in the \texttt{lvcnr} catalog \cite{Schmidt:2017btt}. As a lower cutoff for the match integration, we took the minimum between 20 Hz and the starting frequency of each NR waveform and an upper cutoff at 2048 Hz.
We repeated the calculation for three representative inclinations between the orbital angular momentum and the line of sight $(0, \pi/3, \pi/2)$ and total masses ranging from 50 $M_\odot$ to 350 $M_\odot$.  

\begin{figure}[htpb!]
\includegraphics[width=\columnwidth]{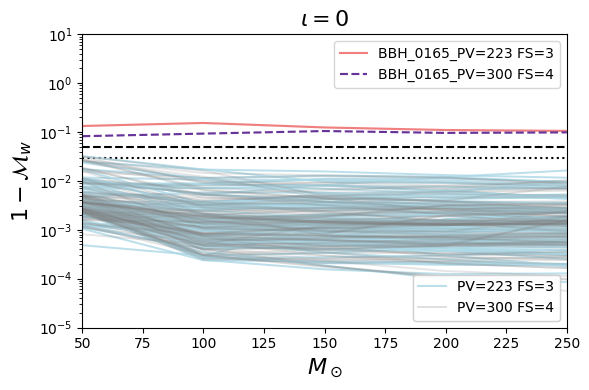}
\includegraphics[width=\columnwidth]{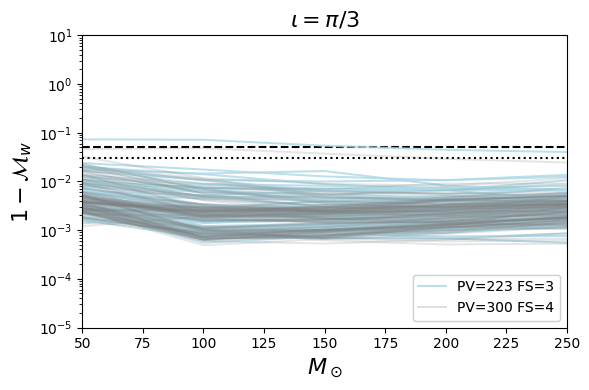}
\includegraphics[width=\columnwidth]{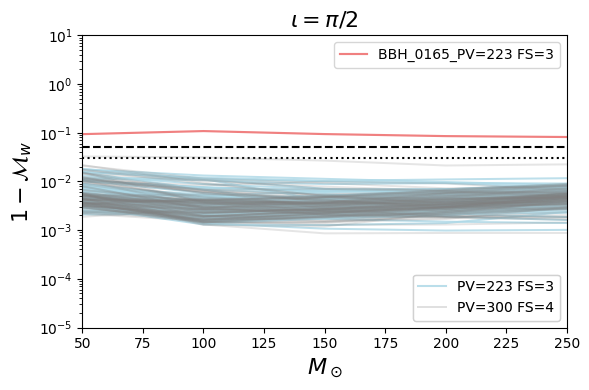}
 \caption{Mismatch comparison against 99 precessing BBH simulations from \texttt{LVCNR} catalog, for different inclinations, as a function of total mass. Dotted black line: 0.03 mismatch. Dashed black line: 0.05 mismatch.}
\label{fig:NRmismatch}
\end{figure}

In Fig. \ref{fig:NRmismatch} we show results for two different versions of the model, $\texttt{PV}=223$ which corresponds to the analytical MSA implementation, and $\texttt{PV}=300$ which is the default numerical evolution implementation explained in Sec.~\ref{sec:Numerical angles}. We observe that the majority of cases have mismatches between $0.001$ and $0.01$, and in general results improve for higher total mass, with a good portion of cases below a mismatch of $0.001$ for face-on inclination. The only outlier in the comparison is \texttt{SXS:BBH:0165}, a very short ($6.5$ orbits) simulation with challenging parameters ($q=6$, $\chi_\mathrm{eff}=-0.43$, $\chi_\mathrm{p}=0.8$). We do observe, however, that the default version of the model considerably improves on the MSA-based twisted-up version (which is more similar to \phXPHM), especially for zero inclination.

\subsubsection{NR injection recovery}

The main target of application of our waveform model is the inference of the source parameters, in particular using Bayesian inference methods to determine the posterior distribution $p({\theta} | {d})$ for the parameters $\theta$ that characterize a binary, given some data $d$. From Bayes' theorem, we have
\begin{align}
    p(\theta | d) &= \frac{\mathcal{L}(d | \theta) \, \pi (\theta)}{\mathcal{Z}} ,
\end{align}
where $\mathcal{L}(d | \theta)$ is the Gaussian noise likelihood \cite{Veitch:2008wd,Veitch:2009hd,Veitch:2014wba}, $\pi (\theta)$ the prior distribution for $\theta$ and $\mathcal{Z}$ the evidence 
\begin{align}
    \mathcal{Z} &= \int d \theta \, \mathcal{L} (d | \theta) \, \pi (\theta) . 
\end{align}
Before testing the model performance on real data from the detectors, it is useful to study how well the model can recover a synthetic signal where we know the parameter values that it should recover. To this end, we have injected synthetic signals into zero noise (i.e., the noise realization corresponding to the average value for Gaussian noise), employing two precessing NR simulations, \texttt{SXS:BBH:0143} and \texttt{SXS:BBH:0062}. As for our studies of mismatches above, to compute the likelihood function we employ the Zero-Detuned-High-Power PSD \cite{adligopsd}. We test different versions of the model and inject the signal at different total mass values (thus changing the number of observable cycles in band). For the analysis here we use  the nested sampling algorithm \texttt{dynesty} \cite{Speagle:2020spe} as implemented in \texttt{Bilby} \cite{Ashton:2018jfp} and \texttt{Parallel Bilby} \cite{Smith_2020}. 

\begin{figure*}[htpb!]
\includegraphics[width=0.5\columnwidth]{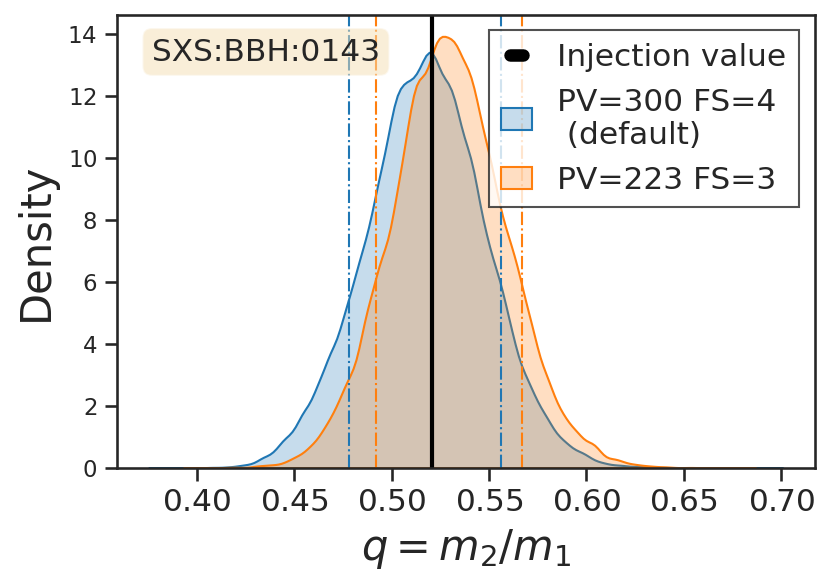}\includegraphics[width=0.5\columnwidth]{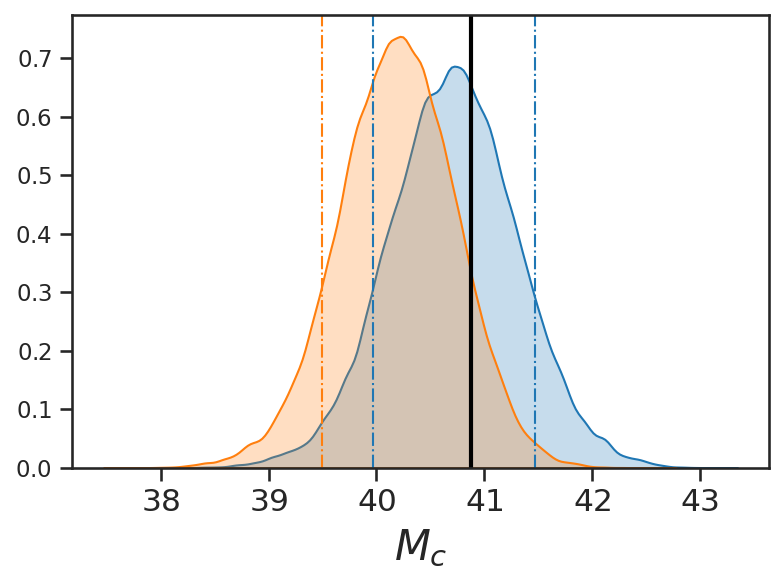}
\includegraphics[width=0.5\columnwidth]{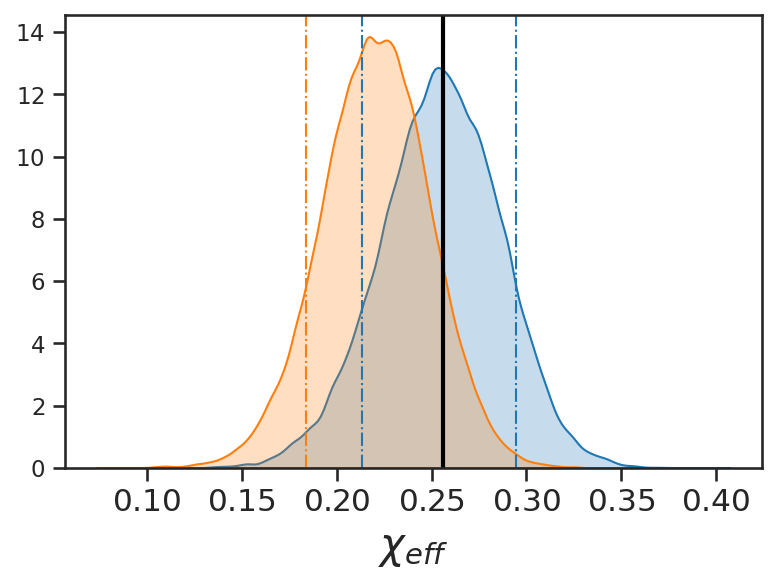}\includegraphics[width=0.5\columnwidth]{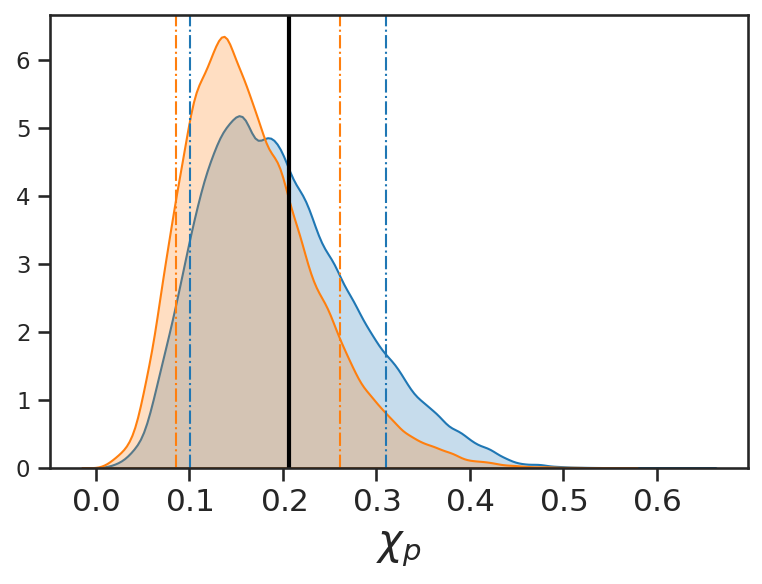}\\
\includegraphics[width=0.5\columnwidth]{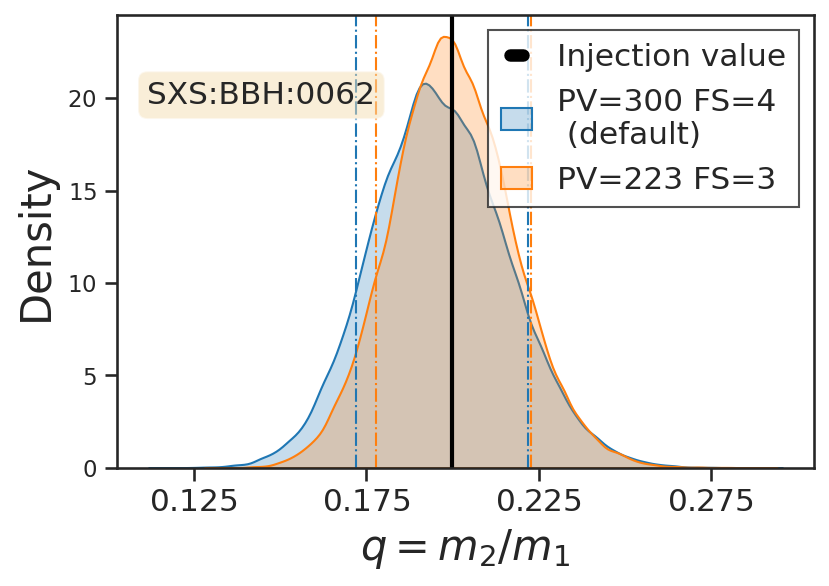}\includegraphics[width=0.5\columnwidth]{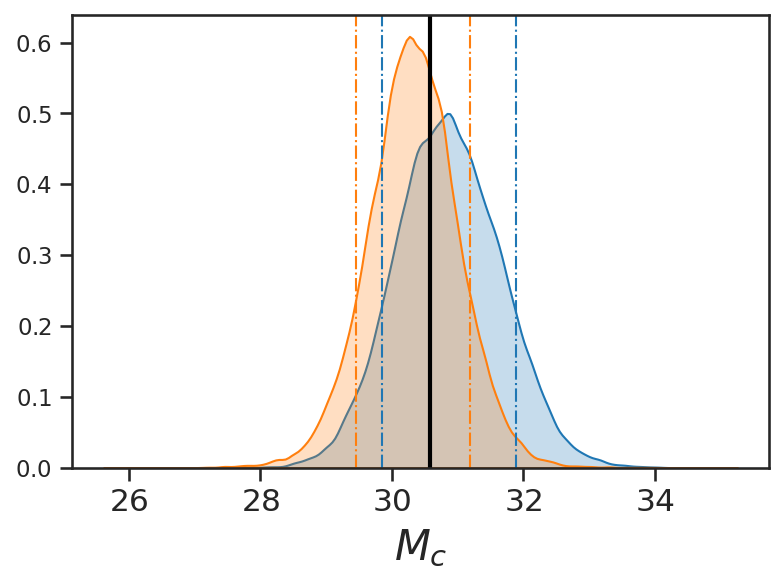}
\includegraphics[width=0.5\columnwidth]{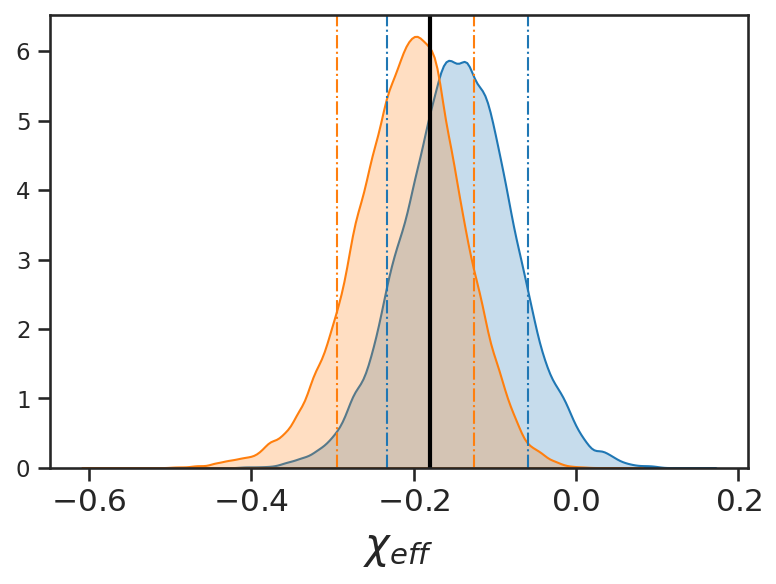}\includegraphics[width=0.5\columnwidth]{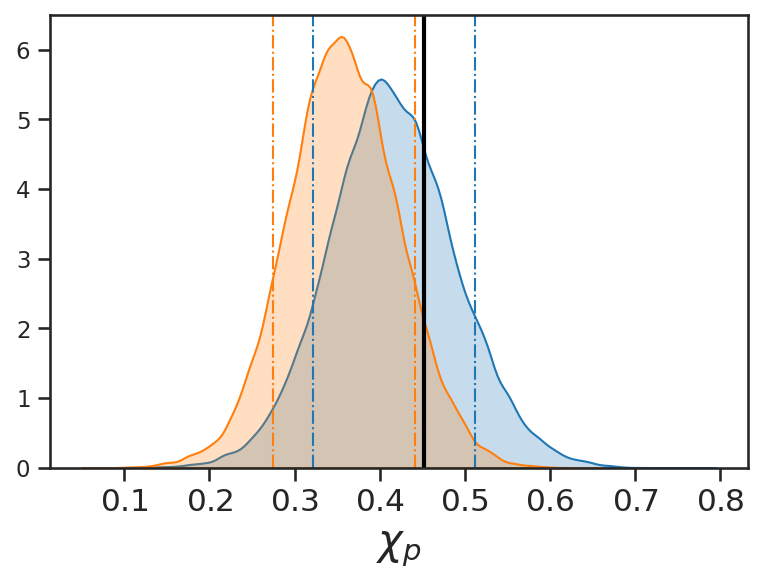}
 \caption{Recovered posterior distributions for the key intrinsic parameters at injected mass $M_T=100M_{\odot}$, comparing the \phTPHM{} standard version ($\texttt{PV=300}$) and the version with MSA Euler angles ($\texttt{PV=223}$). Top: \texttt{SXS:BBH:0143}. Bottom: \texttt{SXS:BBH:0062}. $M_c$ is the chirp mass.}
\label{fig:sxs0143M100}
\end{figure*}

\setlength{\extrarowheight}{7pt}
\begin{table*}[htpb!]
    \centering
    \begin{tabular}{| c | l | l | l | l | l | l |}
\hline
\hline
  & \multicolumn{3}{c|}{\texttt{SXS:BBH:0143}} &\multicolumn{3}{c|}{\texttt{SXS:BBH:0062}} \\ \cline{1-7}
\hline
\hline
$M_\mathrm{tot}/M_\odot$ & 100($99.78^{+1.81}_{-1.77}$) & 200($196.98^{+8.55}_{-8.01}$) & 300($292.66^{+18.0}_{-17.33}$) &  100($101.72^{+5.78}_{-5.1}$) & 200($220.59^{+17.45}_{-17.45}$) & 300($335.31^{+25.33}_{-26.86}$)  \\
 %\hline
 $M_\mathrm{c}/M_\odot$ & 40.88($40.71^{+0.76}_{-0.75}$) & 81.76($79.76^{+4.74}_{-4.79}$) & 122.64($116.78^{+10.0}_{-12.39}$) &  30.58($30.84^{+1.05}_{-0.99}$) & 61.17($82.69^{+11.4}_{-13.95}$) & 91.75($125.57^{+17.22}_{-22.51}$)  \\
 %\hline
 $q$ & 0.52($0.52^{+0.04}_{-0.04})$ & $0.5^{+0.07}_{-0.07}$ & $0.47^{+0.11}_{-0.12}$ & 0.2($0.2^{+0.03}_{-0.02}$) & $0.36^{+0.16}_{-0.12}$ & $0.36^{+0.15}_{-0.12}$ \\
 %\hline
 $\chi_{\text{eff}}$ & 0.25 ($0.26^{+0.04}_{-0.04}$) & $0.23^{+0.08}_{-0.08}$ & $0.24^{+0.13}_{-0.14}$ & -0.18($-0.15^{+0.09}_{-0.09}$) & $-0.21^{+0.16}_{-0.19}$ & $-0.19^{+0.19}_{-0.26}$ \\
 %\hline
 $\chi_\mathrm{p}$  & 0.21 ($0.19^{+0.12}_{-0.08}$) & $0.25^{+0.15}_{-0.12}$ & $0.26^{+0.18}_{-0.14}$ & 0.45($0.41^{+0.1}_{-0.09}$) & $0.37^{+0.26}_{-0.2}$ & $0.33^{+0.33}_{-0.21}$ \\
 %\hline
$\theta_{JN}$ & 0.82 ($0.84^{+0.12}_{-0.12}$) & $0.81^{+0.19}_{-0.18}$ & $0.78^{+0.23}_{-0.22}$ & 0.2($0.28^{+0.13}_{-0.13}$) & $0.52^{+0.54}_{-0.31}$ & $0.53^{+2.01}_{-0.31}$ \\
 %\hline
$\phi_{\text{ref}}$ & 1.5 ($0.95^{+0.43}_{-0.43}$) & $3.89^{+0.45}_{-0.46}$ & $3.31^{+0.52}_{-0.64}$ & 0($0.86^{+5.24}_{-0.7}$) & $3.46^{+0.53}_{-0.56}$ & $1.34^{+4.78}_{-1.19}$ \\
%\hline
$\psi$ & 0.33 ($0.31^{+0.14}_{-0.14}$) & $0.38^{+0.31}_{-0.23}$ & $0.35^{+2.44}_{-0.25}$ & 0.33($1.8^{+0.41}_{-0.41}$) & $1.15^{+0.44}_{-0.48}$ & $1.37^{+0.52}_{-0.47}$ \\
 %\hline
 $\alpha$   & 1.375 ($1.38^{+0.01}_{-0.01}$) & $1.38^{+0.04}_{-0.04}$ & $1.38^{+0.07}_{-0.08}$ & 1.375($1.38^{+0.03}_{-0.04}$) & $1.38^{+0.06}_{-0.08}$ & $1.39^{+0.11}_{-0.12}$ \\
 %\hline
 $\delta$  & -1.21 ($-1.21^{+0.01}_{-0.01}$) & $-1.21^{+0.04}_{-0.03}$ & $-1.21^{+0.07}_{-0.06}$ & -1.21($-1.21^{+0.03}_{-0.03}$) & $-1.21^{+0.06}_{-0.07}$ & $-1.22^{+0.09}_{-0.28}$ \\
 %\hline
 $\rho_\mathrm{mf}^\mathrm{N}$ & 60 ($57.45^{+0.04}_{-0.06}$) & 35 ($33.11^{+0.08}_{-0.11}$) & 25($23.29^{+0.11}_{-0.16}$) & 32($30.0^{+0.08}_{-0.11}$) & 25($23.3^{+0.11}_{-0.16}$) & 25($22.82^{+0.11}_{-0.15}$) \\
 \hline
 \hline
    \end{tabular}
 \caption{Injected parameters  and recovered parameters (median and $90\%$ confidence intervals) for the set of injections employed for testing the model performance. All angle values are given in radians. $\rho_\mathrm{mf}^\mathrm{N}$ is the network signal-to-noise ratio (SNR).}
 \label{tab:injection_settings}
\end{table*}

\begin{figure*}[htpb!]
\includegraphics[width=\columnwidth]{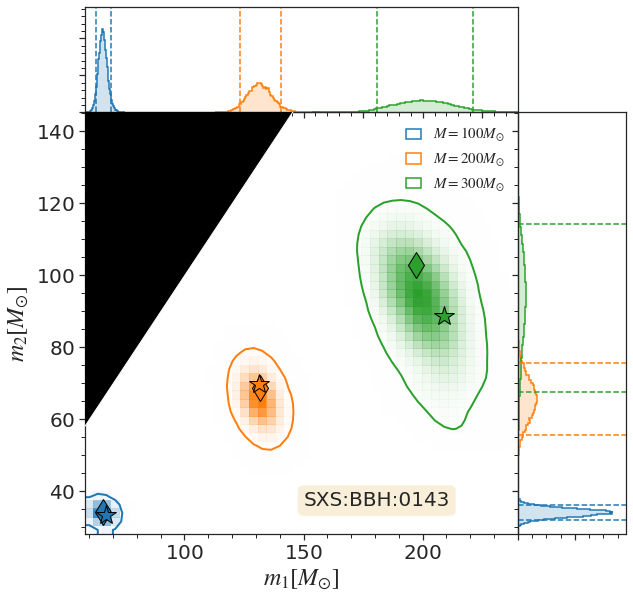}
\includegraphics[width=\columnwidth]{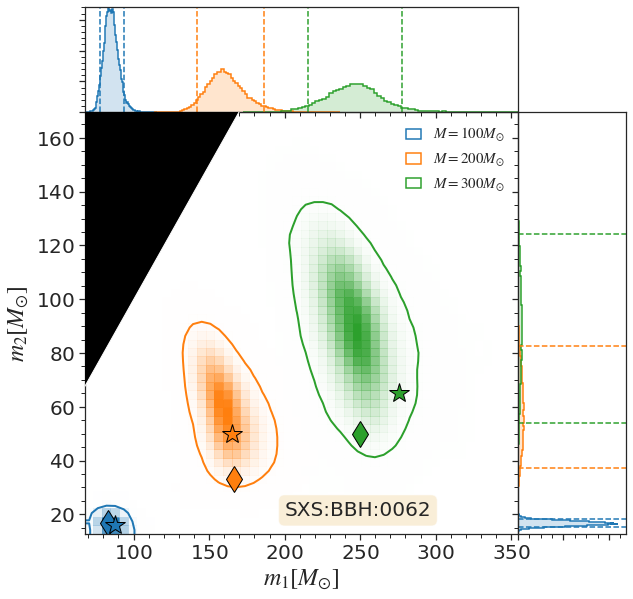}
 \caption{Component mass posterior inference for different total mass injections, recovered with the standard version of \phTPHM{} ($\texttt{PV=300}$). First panel: \texttt{SXS:BBH:0143}. Second panel: \texttt{SXS:BBH:0062}. Diamonds indicate the injected values and stars the maximum likelihood values.}
\label{fig:massseriesNRinj}
\end{figure*}

We have selected fixed extrinsic parameters for each injection. In Table \ref{tab:injection_settings} we list the  parameters for each injection and the recovered median values with $90\%$ confidence interval error estimates. In Fig. \ref{fig:sxs0143M100} we show the recovered posterior distributions for the main intrinsic parameters of both signals injected with a total mass of $100\,M_{\odot}$. We compare the analytical MSA version of the Euler angles ($\texttt{PV=223}$) with the new numerical integration implementation ($\texttt{PV=300}$). For \texttt{SXS:BBH:0143} recovery is better with the default numerical implementation. For \texttt{SXS:BBH:0062} and most parameters the small deviations between the median value of the distribution and the actual injected values are similar for both versions, with however a better recovery of the precessing spin parameter $\chi_\mathrm{p}$ for version $\texttt{PV=300}$.
In Fig. \ref{fig:massseriesNRinj} we show the recovery of the component masses for the different total injected masses, employing the default version of the model. It can be seen that the injected values for the component masses lie inside the $90\%$ confidence intervals for all cases, while there is a degradation of the maximum likelihood prediction as the total mass increases and the observable cycles in the detector are fewer. In general, from the results reported in Table \ref{tab:injection_settings}, it can be seen that the parameter recovery for \texttt{SXS:BBH:0143} is good for all parameters and masses, while for \texttt{SXS:BBH:0062} there is some parameter bias at $M_T=200\,, 300 M_{\odot}$ injections, where several source parameters are not recovered within the 
$90\%$ confidence limits. For the next LIGO-Virgo observation run, O4, where SNR values as large as the ones injected here can be expected, indeed further improvements in the model are foreseen.

One should however be cautious to derive general conclusions about inherent systematic biases from only a few points in the high-dimensional parameter space of CBCs. 
For a given set of intrinsic parameters, a representative subset of extrinsic parameters would have to be selected to perform a systematic analysis, which becomes prohibitively costly with current codes if a significant portion of the intrinsic parameter space has to be studied. Further studies towards understanding waveform modelling caveats for preparing the next, more precise, observing runs of the detectors will require the combination of full parameter estimation with cheaper techniques like fitting factor estimates and Fisher matrix approaches. We leave the development of these studies for future work.
A further issue is that for high total mass, when the available information from the inspiral is smaller and only a few cycles are present in band, degeneracies between the extrinsic parameters and combinations of the intrinsic parameters can complicate the analysis, as we discuss in detail in our re-analysis of the high mass event GW190521 \cite{estelles2021detailed}.

\subsection{Comparison with other state-of-the-art precessing multimode waveform models}\label{sec:modelcomparison}

\begin{figure}[htpb!]
\includegraphics[width=\columnwidth]{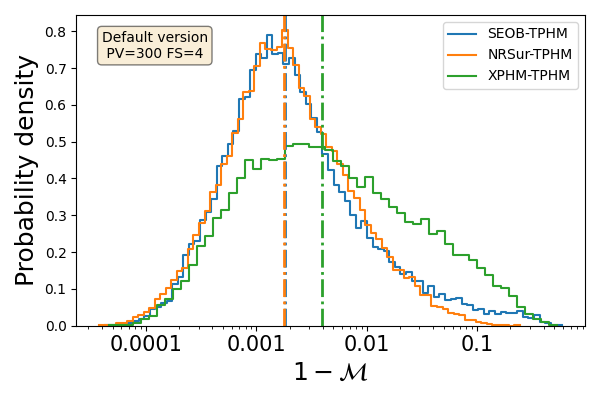}
\includegraphics[width=\columnwidth]{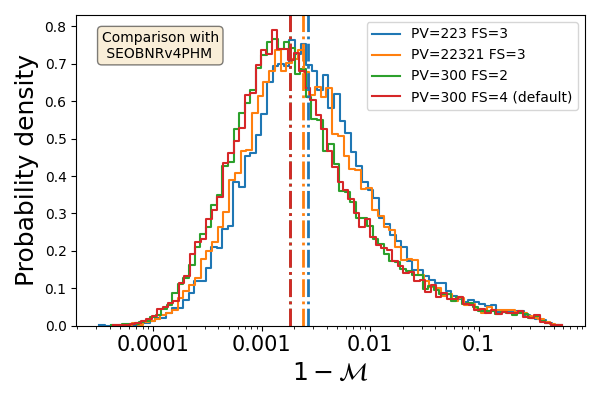}
\includegraphics[width=\columnwidth]{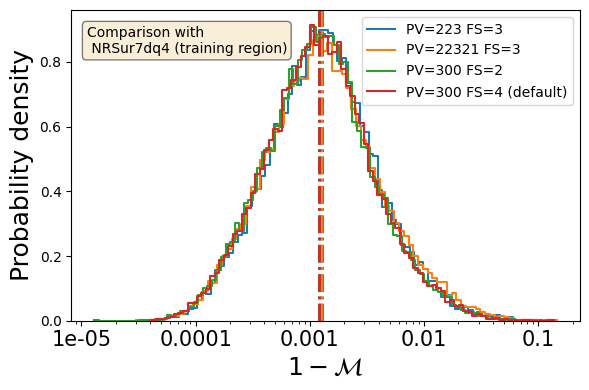}
 \caption{Mismatch comparison with other state-of-the-art precessing multimode waveform models. Top: Mismatch distribution of the default version of \phTPHM{} (numerical evolution of spin equations) against different models (\texttt{SEOBNRv4PHM}: blue, \texttt{NRSur7dq4}: orange, \phXPHM: green). Middle: Mismatch distribution of different versions of the model with respect to \texttt{SEOBNRv4PHM}. Bottom: Mismatch distribution of different versions of the model with respect to \texttt{NRSur7dq4} (in its training region, see main text for more details). Dashed vertical lines: median value of each distribution.}
\label{fig:model300FS4mm}
\end{figure}

\begin{figure}[htpb!]
\includegraphics[width=\columnwidth]{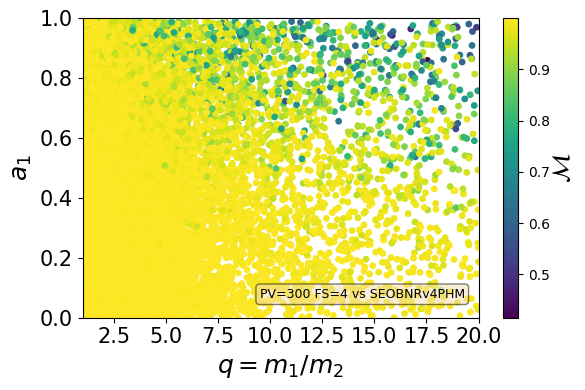}
\includegraphics[width=\columnwidth]{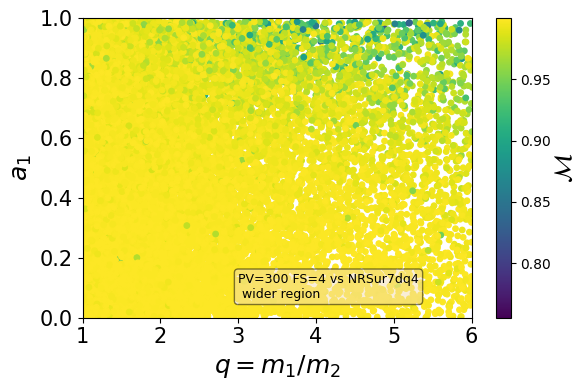}
 \caption{Two-dimensional distribution of matches when comparing the default version of \phTPHM with \texttt{SEOBNRv4PHM} and \texttt{NRSur7dq4}, as a function of the primary spin magnitude and mass ratio. In both cases, matches degrade for very unequal-mass systems with $a_{1}\gtrsim 0.6$.}
\label{fig:scattermm}
\end{figure}

We also compute matches for \phTPHM{} against a number of other state-of-the-art waveform models (\texttt{SEOBNRv4PHM}~\cite{ossokine2020multipolar}, \texttt{NRSur7dq4}~\cite{Varma_2019_prec} and \phXPHM~\cite{phenomxphm}), on two sets of 30000 random configurations, chosen so that the first set lies within the training region for NRSur7dq4 (i.e. $q\leq 4$ and spins isotropically distributed with the constraint $a_{\mathrm{1,2}}\leq 0.8$) while the second set encompasses a wider region of parameter space, allowing mass ratios up to $q=20$ as well as nearly maximally-spinning black holes ($a_{\mathrm{1,2}}\leq 0.99$). We generate the random samples only once and then repeat the match calculation on different pairs of models to ensure a fair comparison. The upper panel of Fig. \ref{fig:model300FS4mm} shows that the bulk of the mismatch distribution between \texttt{IMRPhenomTPHM} and \texttt{SEOBNRv4PHM} or \texttt{NRSur7dq4} lies around $10^{-3}$; the median of the distribution (marked by a dashed line) is shifted towards slightly higher values in the comparison against \phXPHM, which also shows a broader tail towards poor matches. In the middle (bottom) panel of the same figure, we compare the performance of different \texttt{IMRPhenomTPHM} configurations against \texttt{SEOBNRv4PHM} (\texttt{NRSur7dq4}) over the extrapolated (training) region. Versions incorporating a numerical evolution of the spin dynamics (green and red curves) are in closer agreement with \texttt{SEOBNRv4PHM}, while we do not observe significant differences among versions when comparing to \texttt{NRSur7dq4}.
In Fig.~\ref{fig:scattermm}, we show instead the distribution of matches against other time-domain models as a function of the primary spin magnitude and mass ratio. This plot allows to trace the origin of the tail of low matches observed in the preceding histograms: the largest differences between different waveform models are correlated with very unequal masses and highly positive/negative spins. This is expected since, due to the limited availability of numerical waveforms in this region, extrapolation effects are likely to prevail. 

\subsection{Parameter estimation on GW190412}\label{sec:pe}

After analysing the performance of the model compared with precessing waveforms from other state-of-the-art waveform models and NR waveforms, in particular the correct parameter recovery of synthetic injected signals, we now examine the performance of the model analysing a real BBH event, GW190412~\cite{collaboration2020gw190412}, which also has been recently re-analysed with the 4th generation of Phenom waveform models, including the non-precessing version of our model, \texttt{IMRPhenomTHM}, in \cite{colleoni190412} (see also a recent analysis of this event, employing the NR surrogate model \texttt{NRSur7dq4}, as well as the phenomenological models \texttt{IMRPhenomXPHM} and \texttt{IMRPhenomPv3HM}, in \cite{islam2021improved}).

As done in \cite{colleoni190412}, we employ v2 of the strain data~\cite{GW190412:gwosc-v2} for GW190412 released through the Gravitational Wave Open Science Center~(GWOSC) \cite{Vallisneri:2014vxa,Abbott:2019ebz}, with a default sampling rate of 16384\,Hz, for consistency with the official LVC study.
This version has non-linear subtraction~\cite{Vajente:2019ycy} of 60\,Hz power lines applied to it.
We also use the PSDs~\cite{Littenberg:2014oda,Chatziioannou:2019zvs} and calibration uncertainties~\cite{Sun:2020wke} included in v11 of the posterior sample release~\cite{GW190412:dcc} for this event.
We analyse 8\,s of strain data from each of the Hanford, Livingston and Virgo detectors around the trigger time of the event, as reported in GraceDB~\cite{GraceDB}.

\begin{figure*}[htpb!]
\includegraphics[width=0.66\columnwidth]{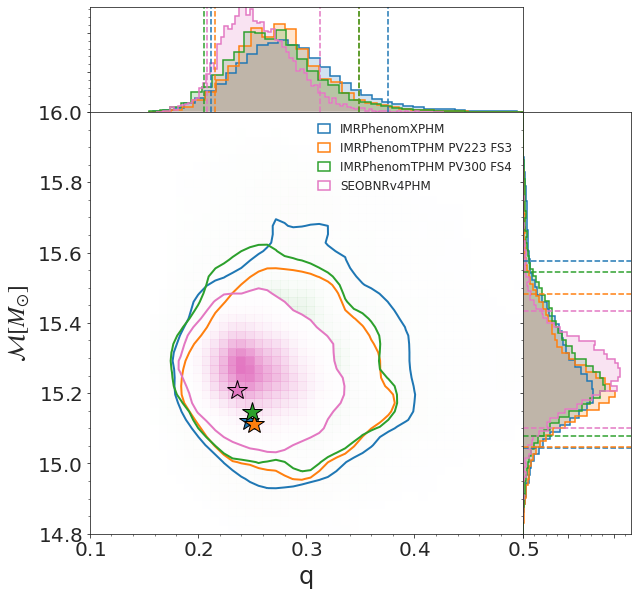}\includegraphics[width=0.66\columnwidth]{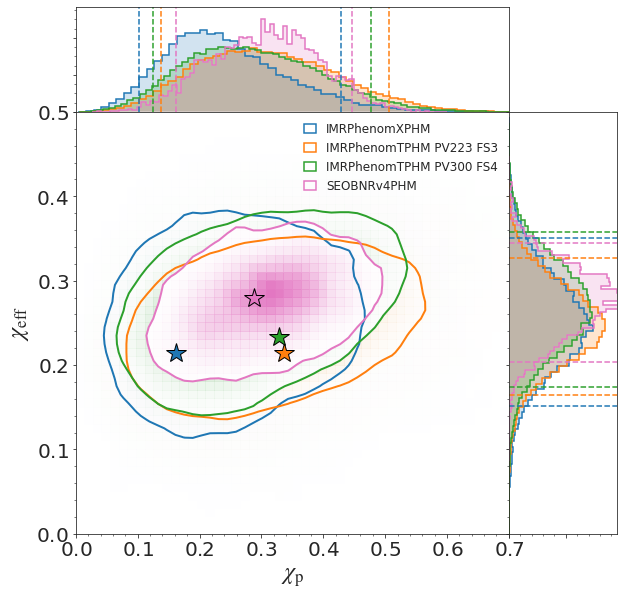}
\includegraphics[width=0.66\columnwidth]{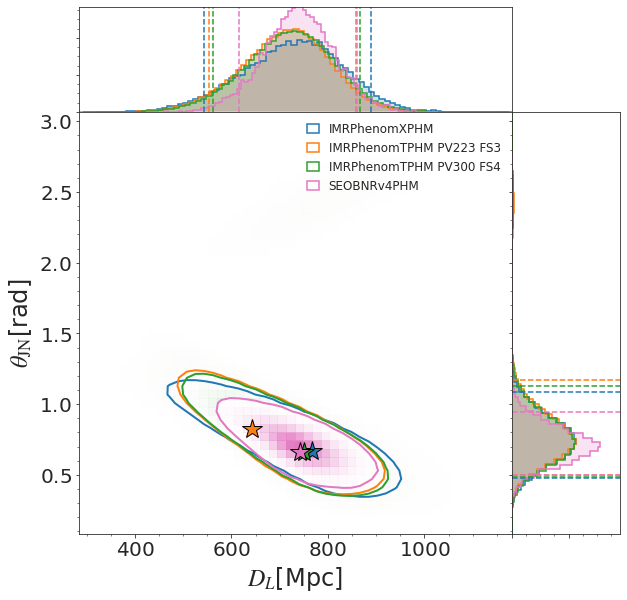}
 \caption{Inferred posterior distributions for the parameters of the BBH event GW190412, comparing our new \phTPHM{} results with those for \texttt{SEOBNRv4PHM} from~\cite{collaboration2020gw190412,GW190412:dcc} and \texttt{IMRPhenomXPHM} from~\cite{colleoni190412}. Stars indicate the maximum likelihood value for each run.}
\label{fig:gw190412corner}
\end{figure*}

In Fig.~\ref{fig:gw190412corner} we compare the results for two versions of \phTPHM{}, analytical MSA angles ($\texttt{PV}=223$) and the default numerical implementation ($\texttt{PV}=300$), with the results for the model \texttt{SEOBNRv4PHM} released with the LVC publication \cite{collaboration2020gw190412,GW190412:dcc} and the preferred results for the standard version of the model \texttt{IMRPhenomXPHM} from the recent re-analysis in \cite{colleoni190412} (we note that the results for \texttt{SEOBNRv4PHM} were obtained with a different parameter estimation pipeline, RIFT \cite{lange2018rapid} and without employing marginalization over detector uncertainty). Recovery of the parameters is quite consistent with the previous published results. Mass ratio is constrained, according to the $90\%$ confidence intervals, to more unequal values than the results for \texttt{IMRPhenomXPHM}, more in accordance with the results from \texttt{SEOBNRv4PHM}. In a similar way, both versions of the model have support for higher values of the effective precessing spin parameter $\chi_\mathrm{p}$, while $\chi_\mathrm{eff}$ is the parameter showing greater difference between both versions: $\texttt{PV=300}$ agrees better with \texttt{SEOBNRv4PHM} and $\texttt{PV=223}$ agrees better with \texttt{IMRPhenomXPHM}. In Table \ref{tab:PEresults0412} we report the median and $90\%$ confidence intervals for the main parameters recovered with version $\texttt{PV=300}$ $\texttt{FS=4}$, showing that all values are consistent with previous published results.

\begin{figure}[htpb!]
\includegraphics[width=\columnwidth]{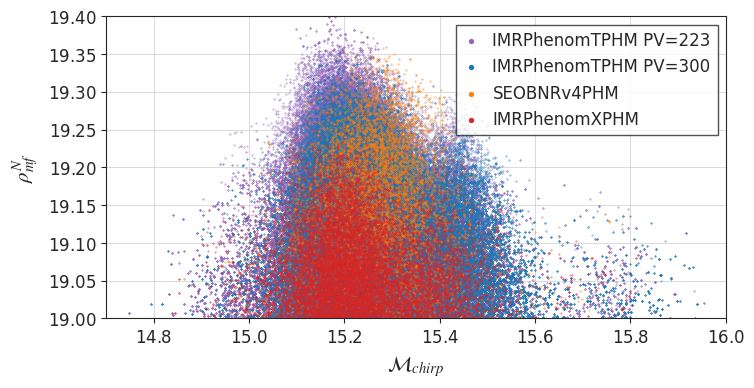}
 \caption{Network matched filter SNR comparison between the different results for the event GW190412.}
\label{fig:gw190412snr}
\end{figure}

\begin{table*}
\caption{Inferred parameter values for GW190412 and
their 90\% credible intervals, obtained using precessing models including higher
multipoles.
Columns 2--4 correspond to the results from the LVC analyses~\cite{collaboration2020gw190412},
the fifth column gives the results from the precessing higher-modes model \phXPHM{} reported in \cite{colleoni190412}, and the last column provides the results obtained with our model, for the default version run.}
\label{tab:PEresults0412}
\begin{ruledtabular}

\begin{tabular}{lrrrrr} 
 parameter
 & \seobnrvforphm & \phPvthreehm & LVC Combined & \phXPHM & \phTPHM \\ \hline
$m_1^\mathrm{s} / M_\odot$ & \MOneSourceCIEOBPHM & \MOneSourceCIPhPHM &
\MOneSourceCICombined  & \MOneSourceCIXPHM & \MOneSourceCITPHM\\
$m_2^\mathrm{s} / M_\odot$ & \MTwoSourceCIEOBPHM & \MTwoSourceCIPhPHM &
\MTwoSourceCICombined  &\MTwoSourceCIXPHM  & \MTwoSourceCITPHM\\
$M^\mathrm{s} / M_\odot$ & \MtotalSourceCIEOBPHM & \MtotalSourceCIPhPHM &
\MtotalSourceCICombined  & \MtotalSourceCIXPHM & \MtotalSourceCITPHM\\
$\mathcal M^\mathrm{s} / M_\odot$ & \ChirpMassSourceCIEOBPHM & \ChirpMassSourceCIPhPHM &
\ChirpMassSourceCICombined   & \ChirpMassSourceCIXPHM & \ChirpMassSourceCITPHM \\
$q$ & \MassRatioCIEOBPHM & \MassRatioCIPhPHM &
\MassRatioCICombined  &\MassRatioCIXPHM & \MassRatioCITPHM \\[6pt]
$\chi_{\rm eff}$ &  \ChiEffCIEOBPHM & \ChiEffCIPhPHM &
\ChiEffCICombined  & \ChiEffCIXPHM & \ChiEffCITPHM\\
$\chi_\mathrm{p}$ &  \ChiPCIEOBPHM & \ChiPCIPhPHM &
\ChiPCICombined  & \ChiPCIXPHM & \ChiPCITPHM \\
$\chi_1$ & \SpinMagOneCIEOBPHM & \SpinMagOneCIPhPHM &
\SpinMagOneCICombined &\SpinMagOneCIXPHM & \SpinMagOneCITPHM
\\[6pt]
$D_\mathrm{L} / \textrm{Mpc}$ &  \DLCIEOBPHM & \DLCIPhPHM &
\DLCICombined & \DLCIXPHM & \DLCITPHM\\
$z$ &  \RedshiftCIEOBPHM & \RedshiftCIPhPHM &
\RedshiftCICombined&\RedshiftCIXPHM & \RedshiftCITPHM  \\
$\hat \theta_{JN}$ &  \ThetaJNCIEOBPHM & \ThetaJNCIPhPHM &
\ThetaJNCICombined & \ThetaJNCIXPHM & \ThetaJNCITPHM \\[6pt]
$\rho_\mathrm{H}$  &  \HSNRCIEOBPHM & \HSNRCIPhPHM &
\HSNRCICombined & \HSNRCIXPHM & \HSNRCITPHM \\
$\rho_\mathrm{L}$  &  \LSNRCIEOBPHM & \LSNRCIPhPHM &
\LSNRCICombined & \LSNRCIXPHM & \LSNRCITPHM \\
$\rho_\mathrm{V}$  &  \VSNRCIEOBPHM & \VSNRCIPhPHM &
\VSNRCICombined & \VSNRCIXPHM & \VSNRCITPHM \\
$\rho_\mathrm{HLV}$  &  \NetSNRCIEOBPHM & \NetSNRCIPhPHM &
\NetSNRCICombined & \NetSNRCIXPHM & \NetSNRCITPHM
\end{tabular}
\end{ruledtabular}
\end{table*}

In Fig. \ref{fig:gw190412snr} we compare the network matched filter SNR for the different models. We can observe that both \texttt{IMRPhenomTPHM} model versions are able to recover higher values of SNR than the equivalent \texttt{IMRPhenomXPHM} run, though the width of the distribution (for the chirp mass in this case) is wider than for \texttt{SEOBNRv4PHM}. In terms of the Bayes factor for the different runs, we can see in Table \ref{tab:tabBF} that both \texttt{IMRPhenomTPHM} versions are slightly preferred with respect to \texttt{IMRPhenomXPHM}, in agreement with the SNR results. In terms of comparison with a non-precessing approximant, we employ the published run for \texttt{IMRPhenomTHM} from \cite{colleoni190412} and we find moderate support for the precessing hypothesis.

\begin{table}
\begin{ruledtabular}
\begin{tabular}{ccc}
 & TPHM vs XPHM & TPHM vs THM  \\
\hline \hline 
TPHM PV=223 & $3.02\pm1.2$ & $2.93\pm1.2$\\
TPHM PV=300 & $2.78\pm1.2$ & $2.69\pm1.2$\\
\end{tabular}
\end{ruledtabular}
\caption{
\label{tab:tabBF}
Comparison of Bayes factor between the different \texttt{IMRPhenomTPHM} runs with the \texttt{IMRPhenomXPHM} results and the non-precessing \texttt{IMRPhenomTHM} results from \cite{colleoni190412}. }
\end{table}

In addition to the analysis of this event, the model has been employed in a re-analysis \cite{mateulucena2021adding} of the more massive events from  the GWTC-1 catalog \cite{Abbott_2019}.
Contrary to our results for GW190412, for several events in our re-analysis results of GWTC-1 the \texttt{IMRPhenomTHM} and \texttt{IMRPhenomTPHM} models recover slightly smaller SNRs and Bayes factors than the corresponding frequency-domain models 
\texttt{IMRPhenomXHM} and \texttt{IMRPhenomXPHM}. The notable exceptions are the most massive event, GW170729, and GW190814, which shows mild support for precession. A likely explanation for this behavior is that, while the description of precession is in general more accurate with the default version of \texttt{IMRPhenomTPHM}, due to the numerical integration of the precession equations and the improved treatment of the merger-ringdown regime, the underlying non-precessing description is in general still less accurate than the \text{IMRPhenomX*} counterpart, especially in the description of the orbital frequency evolution. This trade-off complicates to some extent the analysis of the improvements brought by this new model for smaller masses, and we plan to upgrade the non-precessing description towards the employment of the model in the future planned observation run O4.
We have also re-analysed \cite{estelles2021detailed} the very massive event GW190521 \cite{LIGOScientific:2020stg,Abbott:2020tfl}. In this case we find that as expected \texttt{IMRPhenomTPHM} not only provides a better fit to the data than \texttt{IMRPhenomXPHM}, but also shows a much more consistent behavior when varying the options for precession approximation and final spin (PV and FS).

\subsection{Benchmarks}\label{subsec:benchmarks}

In the previous sections we have tested the accuracy of the model with respect to the predictions of other state-of-the-art waveform models as well as NR simulations, showing also consistent parameter recovery for synthetic and real GW signals. Another important aspect to test is the computational efficiency of the model, since typical parameter estimation runs involve of the order of $10^8$ or more waveform evaluations, so that a model that can be useful for systematic studies of events or for studies of parameter estimation methods has to be computationally efficient.

\begin{figure}[htpb!]
\includegraphics[width=\columnwidth]{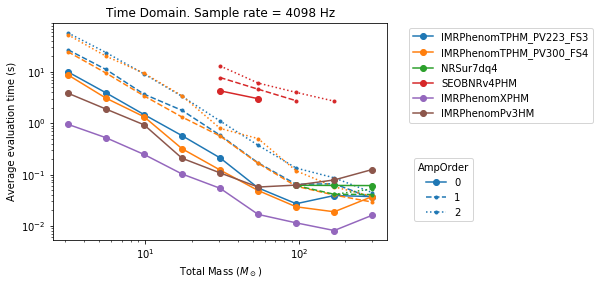}
\includegraphics[width=\columnwidth]{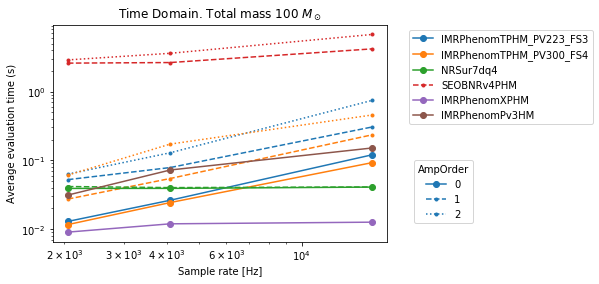}
 \caption{Average evaluation time for the polarizations of different precessing multimode waveform models. Top: Average evaluation time at fixed sampling rate as a function of the total mass of the system. Bottom: Average evaluation time at fixed total mass as a function of the sampling rate.}
\label{fig:benchmarks}
\end{figure}

In Fig. \ref{fig:benchmarks}, we show the average evaluation time for the polarizations, computed as in eq.~(\ref{eq:lalpolariz}), comparing current precessing multimode waveform models. In the top panel, we show the results for a fixed sampling rate of $4096$\,Hz (or, equivalently, a fixed time spacing of $1/4096$\,seconds) varying the total mass of the system (and hence the waveform length) for a fixed bandwidth between 20\,Hz and 2048\,Hz. In the bottom panel, we show the results fixing the total mass of the system to $100\,M_{\odot}$ and varying the sampling rate from 2048\,Hz to 16182\,Hz. Results have been obtained on a Skylake node with a clock speed of 2401\,MHz of the CIT cluster. From this, we can extract the conclusion than for masses greater than $50\,M_{\odot}$ and for typical sampling rates at these masses (2048\,Hz, 4096\,Hz), the \phTPHM{} model has the second fastest evaluation time from the analysed set, only after \texttt{IMRPhenomXPHM}, where extra optimizations have been applied, such as the implementation of the multibanding technique \cite{Garcia-Quiros:2020qlt}. 

We also show results for lowering the starting waveform generation frequency, regulated by the \texttt{amp-order} parameter:
\begin{equation}
    f_\mathrm{start}=\frac{2f_{\min}}{2+\texttt{amp-order}},
\end{equation}
where $f_\mathrm{start}$ is the frequency for starting the waveform generation and $f_{\min}$ is the starting frequency for the noise-weighted inner product in eq.~(\ref{eq:nwip}). This is required for consistently including all the subdominant mode content in band once the waveforms are Fourier-transformed to compute noise-weighted inner products, due to the frequency scaling $\dot{\phi}_{lm}\approx(m/2)\dot{\phi}_{22}$ of the subdominant harmonics frequencies. For example, in order to have all modes in band up to $m=4$ at $f_{\min}$, one needs to set $\texttt{amp-order}=2$. This translates into an increased waveform length. While we can see that increasing the waveform length has a big impact at low masses, we can see that differences are reduced at high masses, which is the preferred applicability regime of this model.

Besides waveform evaluation time, it is useful to examine the mean likelihood evaluation time, since this is a key value for parameter estimation applications: on top of the waveform evaluation cost, this number incorporates the cost of conditioning and Fourier-transforming the polarizations. In Table \ref{tab:tabBench} we can see the mean likelihood evaluation time for several models, split into different mass bins and for different durations of the analyzed data segments. We can see that while at low masses (from $10\,M_{\odot}$ to $60\,M_{\odot}$) the model is five times more expensive than the reference model \text{IMRPhenomXPHM} (the same being true for the comparison of dominant-mode models between \text{IMRPhenomTP} and \text{IMRPhenomXP}), at high mass the difference is approximately only a factor 2, which is quite remarkable taking into account that \text{IMRPhenomTPHM} needs a numerical Fourier transform before being applied in the likelihood evaluation. Noticeably, the model is almost two orders of magnitude faster than the time-domain model \texttt{SEOBNRv4PHM}.

\begin{table*}[htpb]
\begin{center}
 \def\arraystretch{1.3 }
\begin{tabular}{  c  c  c  c  c  c c   c   c  }
\hline
\hline
$[M_{\min},M_{\max}]$ &  $\Delta T$ & \fontsize{8}{104}\selectfont IMRPhenomTP & \fontsize{8}{104}\selectfont IMRPhenomXP & \fontsize{8}{104}\selectfont  SEOBNRv4P & \fontsize{8}{104}\selectfont IMRPhenomTPHM & \fontsize{8}{104}\selectfont IMRPhenomXPHM & \fontsize{8}{104}\selectfont SEOBNRv4PHM & \fontsize{8}{104}\selectfont NRSur7dq4   \\
\hline
\multirow{2}{*}{[60,100]} & 4\,s     &  15.0   & 7.0     &    2006.0  &  55.0  & 24.2    &   4563.2   &  30.4 \\
                          & 8\,s     &   14.7   &  13.9   &   1968.2   &  55.8   & 39.6    &  4717.7   & 30.7 \\
\hline
\multirow{2}{*}{[10,60]}& 4\,s     &    73.9  &  14.7    & 4614.6  &   271.6   &    54.2   &  4729.6   & - \\
                         & 8\,s     &    75.5  &   29.2  &  4672.2  &  267.3  &     84.8   &  5051.3  & - \\
\hline
\hline
 \end{tabular}
\end{center}
\captionof{table}{%\toni{Numbers not final yet}.
Mean likelihood evaluation time in milliseconds for several precessing models including higher-order modes for equal-mass signals. The numbers represent averages over two different total mass ranges $[M_{\min},M_{\max}]=\{[10,60],[60,100]\}M_\odot$ and random spin orientations and magnitudes. The first column indicates the total mass range in which the models are evaluated and the second one specifies the data analysis segment length in seconds used for the calculations.}
\label{tab:tabBench}
\end{table*}

\section{Conclusions}\label{sec:conclusions}

In this paper we have presented a precessing time domain model for the GW signal of coalescing black holes,
\text{IMRPhenomTPHM}, which can be considered to be a time-domain companion to the \text{IMRPhenomXPHM} frequency-domain model. Working in the time domain allows several improvements over the accuracy of \text{IMRPhenomXPHM}: The inspiral description is more accurate thanks to the numerical integration of the post-Newtonian spin evolution in eq. (\ref{eq:preceqs}). The merger is improved by only modifying the non-precessing waveforms, which we take from our \text{IMRPhenomTHM} model, in the ringdown to adapt to the precessing final spin value. In the time domain this is straightforward by only modifying the ringdown portion of the waveform, however this cannot be cleanly translated to the frequency domain, due to the ``smearing'' effect of the Fourier transform. Furthermore we have included an additional option to the analytical description of the Euler angles to smooth the behavior at merger, substituting our ignorance about the actual plunge dynamics by a well-behaved simpler description, from the MECO time to the coalescence time. Finally, in the ringdown we can ensure a consistent behavior with black-hole-perturbation theory by determining the Euler angles from the QNM frequencies. Our approach opens up several avenues for further improvements of the model. One natural avenue is full calibration to numerical relativity precessing simulations, which we expect to be simpler in the time domain. Another new possibility is to analytically tune the information that enters the numerical evolution of the spin equations (\ref{eq:preceqs}), e.g. by incorporating the in-plane spin components in the phasing of the co-precessing modes.

The main application that we foresee for this new model in GW data analysis are high-mass events, where the additional accuracy in treating precession can play a crucial role in the recovery of the source parameters. Indeed, in two accompanying papers \cite{mateulucena2021adding,estelles2021detailed} we find that for the two high-mass events GW170729 and in particular GW190521 \text{IMRPhenomTPHM} matches the data better than \text{IMRPhenomXPHM}, and should be considered as improving over the observational results obtained with \text{IMRPhenomXPHM}.

In the future we plan to improve both the frequency-domain and time-domain \text{IMRPhenom} models, and calibrate them to precessing NR simulations. Beyond quasi-circular systems we also expect that the development of eccentric waveform models, and in particular precessing eccentric ones, will benefit from the insights gained both from the frequency and time domain strategies. Finally, we note that part of the original motivation for time-domain models was to serve as an alternative baseline to develop tests of general relativity, such as inspiral-merger-ringdown or parameterized tests \cite{Ghosh:2017gfp,LIGOScientific:2019fpa,Abbott:2020jks}. To this end, different phenomenological parameterizations of frequency and time-domain models will help to 
assess the robustness of such tests. Moreover, we note that a time-domain treatment guarantees a cleaner separation between the ringdown and the inspiral/merger regimes, while this is not possible in the frequency domain due to smearing effect of the Fourier transform.

\section*{Acknowledgements}
We thank the reviewers of the LIGO Scientific Collaboration, Maria Haney, Jonathan Thompson, Eleanor Hamilton and Jacob Lange for reviewing the \texttt{LALSuite} code implementation, and for carefully reading the manuscript and valuable feedback.

This work was supported by European Union FEDER funds, the Ministry of Science, 
Innovation and Universities and the Spanish Agencia Estatal de Investigación grants
PID2019-106416GB-I00/AEI/10.13039/501100011033,  % current edition of national "rolling grant"
FPA2016-76821-P,     % previous national grant, and for reference older grants: FPA2017-90687-REDC, FPA2017-90566-REDC. 
RED2018-102661-T,    % RENATA
RED2018-102573-E,    % REDES ESTRATÉGICAS: Participación Española en Estructuras Euro... 
FPA2017-90687-REDC,  % CPAN
% Check: removed in Rodrigo's 2020 methods paper
Vicepresidència i Conselleria d’Innovació, Recerca i Turisme, 
% This is Maite's grant
Conselleria d’Educació, i Universitats del Govern de les Illes Balears i Fons Social Europeu, 
Comunitat Autonoma de les Illes Balears through the Direcció General de Política Universitaria i Recerca with funds from the Tourist Stay Tax Law ITS 2017-006 (PRD2018/24),
Generalitat Valenciana (PROMETEO/2019/071),  
EU COST Actions CA18108, CA17137, CA16214, and CA16104, and
the Spanish Ministry of Education, Culture and Sport grants FPU15/03344 and FPU15/01319.
M.C.~acknowledges funding from the European Union's Horizon 2020 research and innovation programme, under the Marie Skłodowska-Curie grant agreement No. 751492.
D.K.~is supported by the Spanish Ministerio de Ciencia, Innovaci{\'o}n y
Universidades (ref.~BEAGAL 18/00148)
and cofinanced by the Universitat de les Illes Balears.
The authors thankfully acknowledge the computer resources at MareNostrum and the technical support provided by Barcelona Supercomputing Center (BSC) through Grants 
No. 
AECT-2020-2-0015,  % The end of an era - analysing the last gravitational wave detections before LIGO-Virgo design sensitivity
AECT-2020-1-0025,  % Testing models for gravitational waves from coalescing black holes with generic configurations
AECT-2019-3-0020,  % Testing models for gravitational waves from coalescing black holes with generic configurations
AECT-2019-2-0010,  % Highly accurate generic black-hole binary simulations: exploring the highly eccentric precessing case
AECT-2019-2-0017,  % Modelling a competition of mass ratio and spin terms for gravitational waves from coalescing black holes
AECT-2019-1-0022,  % Highly accurate gravitational wave signals from spinning eccentric black hole mergers 
from the Red Española de Supercomputación (RES).

This material is based upon work supported by NSF’s LIGO Laboratory which is a major facility fully funded by the National Science Foundation.
The authors also acknowledge the computational resources at the cluster CIT provided by LIGO Laboratory and supported by National Science Foundation Grants PHY-0757058 and PHY-0823459.
This research has made use of data obtained from the Gravitational Wave Open Science Center~\cite{GWOSC}, a service of LIGO Laboratory, the LIGO Scientific Collaboration and the Virgo Collaboration. LIGO is funded by the U.S. National Science Foundation. Virgo is funded by the French Centre National de Recherche Scientifique (CNRS), the Italian Istituto Nazionale della Fisica Nucleare (INFN) and the Dutch Nikhef, with contributions by Polish and Hungarian institutes.

\appendix

\section{Usage of the LALSimulation implementation}\label{appen:useropts}

The \phTPHM{} model has been implemented as C code in the \texttt{LALSimulation} package of the \texttt{LALSuite} \cite{lalsuite} software framework for GW data analysis. Time-domain polarizations can be called through the standard interface \texttt{SimInspiralChooseTDWaveform}. Also spherical harmonic modes in the $\boldsymbol{L}_0$-frame satisfying LAL conventions can be called through the \texttt{LALSimulation} \texttt{SimInspiralChooseTDModes} function. 

For selecting the specific description of the precessing Euler angles detailed in Sec. \ref{sec:EulerAnglesInsp} and \ref{sec:EulerAnglesRD}, the user can specify the parameters \texttt{PhenomXPrecVersion} (\texttt{PV}), in a three or five digit format. The first three digits correspond to the core specification of the Euler angles description: 102 for NNLO, 223 for MSA and 300 for numerical evolution of the spin precession equations (alternative versions of NNLO and MSA can also be specified, see Table III of Appendix F in \cite{phenomxphm}). For the analytical descriptions, the fourth digit selects the merger-ringdown treatment: 0 corresponds to disabling the ringdown angle approximation from eq.~(\ref{eq:anglesrd}), 1 to enabling it, and 2 to enabling it with the addition of the linear continuation from MECO time to coalescence time in eq.~(\ref{eq:linearmerger}). The fifth digit corresponds to the treatment of the third Euler angle: 0 corresponds to the analytical expression and 1 to numerical evaluation. Our default implementation with $\texttt{PV}=300$ does not include additional options, and it always enables the ringdown angles approximation described in eq.~(\ref{eq:anglesrd}). See Table \ref{tab:prec_version} for a summary of available options.

The final spin description can be selected via the parameter \texttt{PhenomXFinalSpinMod} (\texttt{FS}). Available options inherited from the \texttt{IMRPhenomXP/PHM} implementation are described in Table V of Appendix F in \cite{phenomxphm}. In this model we have incorporated a new default option, described in Sec.~\ref{sec:EulerAnglesRD}, where the individual spins are evaluated at coalescence time from the evolution of eq.~(\ref{eq:preceqs}). This option is the default and can also be explicitly specified with option $\texttt{FS}=4$, but it is only available with $\texttt{PV}=300$ since it requires the numerical evolution of the individual spins.

\begin{table*}[htpb]
\centering
    \begin{tabular}{ | l | l | }
\hline
\texttt{PV} & Explanation \\ \hline\hline
10200/22300    &  NNLO/MSA analytical expressions evaluated for the whole waveform.\\ \hline
10210/22310     &  NNLO/MSA analytical expressions evaluated until time $t_\mathrm{c}$.\\
& Attach ringdown angles with approximation from eq.~(\ref{eq:preceqs}).\\ \hline
10211/22311     &  NNLO/MSA analytical expressions evaluated until time $t_\mathrm{c}$. Attach ringdown angles with approximation from eq.~(\ref{eq:preceqs}).\\
& Euler angle $\gamma$ substituted by numerical evaluation of the minimal rotation condition.\\ \hline
10221/22321     &  NNLO/MSA analytical expressions evaluated until the MECO time. Linear continuation performed until time $t_c$.\\
& Attach ringdown angles with approximation of eqs. (\ref{eq:preceqs}). Euler angle $\gamma$ computed numerically.\\ \hline
102/223     &  Equivalent to 10210/22310. \\ \hline
300     &  Numerical evolution of $\hat{\boldsymbol{L}}(t)$ for obtaining Euler angles until time $t_\mathrm{c}$.\\
& Attach ringdown angles approximation of eqs. (\ref{eq:preceqs}).\\ \hline
\end{tabular}
    \caption{Options in the LALSuite implementation to change between different descriptions of the Euler angles. The coalescence time is denoted by $t_\mathrm{c}$. }
    \label{tab:prec_version}
\end{table*}

\bibliography{ligo.bib, pn.bib, ringdown.bib, wfmodels.bib, misc.bib, pe.bib, prec.bib}{}

\end{document}